\documentclass[pdflatex,sn-mathphys-num,iicol]{sn-jnl}

\usepackage{graphicx}%
\usepackage{multirow}%
\usepackage{multicol}
\usepackage{amsmath,amssymb,amsfonts}%
\usepackage{amsthm}%
\usepackage{mathrsfs}%
\usepackage[title]{appendix}%
\usepackage{xcolor}%
\usepackage{textcomp}%
\usepackage{manyfoot}%
\usepackage{booktabs}%
\usepackage{dcolumn}
\usepackage{latexsym}
\usepackage{bbm}
\usepackage{physics}
\usepackage{caption}
\usepackage{subcaption}
\usepackage{slashed}
\usepackage{ulem}
\usepackage{float}

\theoremstyle{thmstyleone}%
\theoremstyle{thmstyletwo}%
\theoremstyle{thmstylethree}%

\begin{document}

\title{Resurgence in scalar and spinor QED: The Euler-Heisenberg Lagrangian in parallel field backgrounds}

\author[1,2]{\fnm{Drishti} \sur{Gupta}}\email{dg36@illinois.edu}

\author[1]{\fnm{Arun M.} \sur{Thalapillil}}\email{thalapillil@iiserpune.ac.in}

\affil[1]{\orgdiv{Department of Physics}, \orgname{Indian Institute of Science Education and Research Pune}, \orgaddress{\street{Homi Bhabha Road, Pashan}, \city{Pune}, \postcode{411008}, \country{Pune}}}

\affil[2]{\orgdiv{Department of Physics}, \orgname{University of Illinois Urbana-Champaign}, \orgaddress{\city{Urbana}, \state{IL}, \postcode{61801}}}

\abstract{We present the first systematic resurgent analysis of the Euler--Heisenberg Lagrangian in spinor and scalar quantum electrodynamics for the most general constant background field configuration. In contrast to the extensively studied case of the perpendicular electric and magnetic fields configuration, the parallel fields configuration exhibits unique asymptotic structures, leading to a substantially richer pattern of singularities in the Borel plane. Explicit large-order asymptotic formulas for the weak-field coefficients in both spinor and scalar quantum electrodynamics are derived. These reveal a nontrivial interplay between alternating and non-alternating factorial growth, governed by distinct structures associated with electric and magnetic contributions, and smoothly interpolating between the known perpendicular fields limit. Using Borel--dispersion techniques, we demonstrate that the complete instanton structure underlying Schwinger pair production in parallel fields backgrounds is encoded in the divergent perturbative coefficients. We then construct resurgent approximants using Pad\'{e}--Borel and Pad\'{e}--Conformal--Borel resummation schemes adapted to the parallel fields configuration. For the spinor case, conformal improvement results in a significant enhancement in reconstructing both the real and imaginary parts of the effective Lagrangian across a wide range of field ratios. Detailed comparisons with exact special-function representations demonstrate the reliability of reconstructions from a modest number of weak-field coefficients. This work establishes a natural completion of the resurgence programme for constant electromagnetic backgrounds, providing a robust analytic framework for exploring nonperturbative physics and strong-field phenomena in spinor and scalar quantum electrodynamics, from finite perturbative data.}


\maketitle

\section{Introduction}

The Euler–Heisenberg Lagrangian (EHL) plays a central role in quantum electrodynamics (QED), serving as the universal effective action obtained by integrating out charged matter in a constant electromagnetic field background.
\begin{equation}
    \mathcal{L}_{\text{HE}} = - \frac{1}{4}F^{\mu \nu} F_{\mu \nu} +  \mathcal{L}^{(1)}+\mathcal{L}^{(2)}+\ldots \; .
\end{equation}
Here, the superscripts denote the loop order. Since the foundational works of Euler, Heisenberg, and Weisskopf\,\cite{Heisenberg-Euler-OG-Paper,Weisskopf1936}, and their later proper–time formulation by Schwinger\,\cite{Schwinger-Paper}, the EHL has served as a framework for exploring various nonlinear and nonperturbative phenomena in quantum electrodynamics (for instance, see\,\cite{Dunne:2012vv} and references therein). Even in recent years, it has continued to catalyse developments in various areas---for instance, in mathematical physics\,\cite{Murcia:2025psi,Russo:2024llm,Sorge:2024gag} and in elucidating structural aspects of effective field theories\,\cite{Gies:2016yaa,Fuentes-Martin:2024agf,Santos:2023ooc,Karbstein2024}.
 
This setting has also emerged as a fertile ground for the modern program of resurgence and resummation theory in physics (see, for instance, \cite{ecalle1985fonctions, Costin2008,DORIGONI2019167914, Aniceto:2018bis, Dunne:2025mye}), which, in the present context of interest, seeks to decode the full nonperturbative content of quantum field theories (see, for example,\,\cite{Polyakov1987,Coleman1985Aspects,Witten:1997fz} and associated references) from asymptotic perturbative data alone\,\cite{dyson, Bender:1969si, Costin:2019xql,Marino:2023epd}. 

Since its introduction, resurgent analysis has been applied across a broad range of physical and mathematical settings; see, for example, \cite{Voros1983, Jentschura:2004jg, Dunne:2012ae, Argyres:2012ka, Dunne:2012zk, Marino:2012zq,Aniceto:2014hoa,Dunne:2016nmc}. In recent years, resurgence has remained an especially active area of investigation. To give a representative, albeit non-exhaustive, sampling of recent developments--- resurgence has been explored in studies of large-order behaviour in quantum chromodynamics \cite{MiravitllasMas:2019neb}, resurgent extrapolation using the Painlev\'e~I equation as a prototype \cite{Costin:2019xql}, superconductivity \cite{Marino:2019wra}, renormalisation group equations \cite{Bersini:2019axn}, applications to the one-dimensional Hubbard model \cite{Marino:2020dgc}, quantum metrics in anharmonic oscillators\,\cite{Hernandez:2025xxe}, quantum knot invariants \cite{Garoufalidis:2020nut}, renormalons in six-dimensional $\phi^3$ theories \cite{Borinsky:2022knn}, investigations of the $O(3)$ and $O(4)$ sigma models \cite{Bajnok:2021zjm,Bajnok:2021dri}, analyses of integrable quantum field theories \cite{DiPietro:2021yxb}, large-charge expansions in $O(2N)$ models \cite{Dondi:2021buw}, studies of the $SU(2)$ Chern--Simons partition function \cite{Wu:2020dhl}, investigations of $sl(2,\mathbb{C})$ Chern--Simons state-integral models \cite{Duan:2022ryd}, the explicit construction of four-dimensional de~Sitter space via nodal diagrams arising from Glauber--Sudarshan states \cite{Brahma:2022wdl}, applications in matrix models and D-brane physics \cite{Schiappa:2023ned}, analyses of conformal blocks in the large central-charge limit \cite{Benjamin:2023uib}, studies of supersymmetric $\mathcal{N}=1$ Jackiw--Teitelboim gravity \cite{Griguolo:2023jyy}, resurgence in the Nambu-Jona-Lasinio model\,\cite{Bersini:2025lxs}, in B-model topological string theories\,\cite{Li:2025zyr} and investigations at the interface of quantum topology and mathematical physics \cite{Costin:2023kla}, among many others. Even this partial sampling of studies is a testament to the breadth and potential of these concepts across a wide range of problems in theoretical and mathematical physics.

In the context of spinor and scalar quantum electrodynamics (QED and SQED, henceforth), recent results from the application of these methods has illuminated various aspects of the resurgence and resummation programme, as when it is applied to real-world quantum field theoretic settings\,\cite{Dunne:1999uy,florio,DunneHarrisInhom,dunne-harris-higher-loop,Gupta:2023gsw,non-linear-trident-resummation,Torgrimsson:2022ndq,Torgrimsson:2021wcj,Torgrimsson:2021zob}. Among other things, QED and SQED, being relatively well-understood quantum field theories, give a unique opportunity to test various aspects of the reconstruction, thereby effectively acting as a validation framework and dictionary. Apart from these, there are also other compelling reasons. The most general higher-loop, weak-field instanton expansion of the nonperturbative imaginary part of the effective Lagrangian is unknown\,\cite{Huet:2017}, with even today the full two-loop order expansion only having a conjectured form\,\cite{Ritus:1975,Ritus:1977a,LebedevRitus:1984}. Additionally, in many applications, attempting to construct exact closed-form representations and leveraging them for computing realistic observables appears overly tedious, even for $1+1$ QED at just three-loop orders\,\cite{Huet:2009,Huet:2017,Huet:2019}. These endeavors have gained even more relevance with the advent of various high-intensity light sources globally where strong-field techniques and phenomena abound (see, for instance,\,\cite{Fedotov_2023, Kropf:2025loq} and related references). In addition, strong-field phenomena are also ubiquitous in various astrophysical and cosmological settings\,\cite{Ruffini:2009hg,Kim:2019joy}. Thus, there are compelling reasons for attempting to understand and reconstruct nonperturbative information in QED and SQED from finite-order perturbative data, using resurgent and resummation techniques.

For homogeneous electromagnetic backgrounds with perpendicular electric and magnetic fields, Padé–Borel reconstructions have been shown to recover the imaginary part of the QED Lagrangian with remarkable precision\,\cite{florio}. Inhomogeneous-field studies\,\cite{DunneHarrisInhom} in QED demonstrate that resurgent asymptotics not only capture the factorial growth of weak-field coefficients but also encode additional branch-point structures absent in homogeneous systems. It has also been shown that the two-loop (and higher-loop) Euler–Heisenberg QED effective Lagrangian exhibits a qualitatively new resurgent trans-series structure, in which the simple Borel poles present at one loop are replaced by branch-point singularities due to interactions among virtual particles\,\cite{dunne-harris-higher-loop}. Resummation methods have also been applied to the nonlinear trident process and radiation reaction in intense laser fields\,\cite{non-linear-trident-resummation,Torgrimsson:2022ndq,Torgrimsson:2021wcj,Torgrimsson:2021zob}. Our earlier work on the resurgence structure of the SQED EHL\,\cite{Gupta:2023gsw} illustrated how it contrasts with that in QED and how the one-loop and two-loop asymptotics conspire to reproduce strong-field behaviour and clarified the interplay between one-particle-reducible and one-particle-irreducible contributions.

To date, most resurgent analyses of the Euler–Heisenberg effective action have focused on perpendicular field configurations. Because these backgrounds can be Lorentz-boosted to a frame with a single non-zero field, they simplify to purely magnetic or purely electric limits; in the latter case, the emergence of an imaginary part at strong fields physically manifests as Schwinger pair production. In these single-field cases, the weak-field series exhibits a factorial divergence with a large-order structure indicative of a $\Gamma(2n+2)$ growth at one-loop order, leading to a tower of Instanton contributions 
\begin{equation}
I_k=C_k \exp \left(-\frac{k\pi m^2}{qE} \right) \qquad (k=1,2,\ldots)\; .
\end{equation}
where $C_k$ is a prefactor multiplying instanton contributions. This structure underlies the existence of a Borel transform with poles at $s=\pm i k\pi$, enabling Pad\'{e}–Borel techniques to reconstruct both the real and imaginary parts of the QED and SQED effective actions with surprising accuracy from weak-field data\,\cite{florio,Gupta:2023gsw}. At higher-loop order, the transseries structure is much richer\,\cite{dunne-harris-higher-loop,Gupta:2023gsw} and already at two-loop order, the complete weak-field expansion of the nonperturbative imaginary part is only a conjecture\,\cite{Ritus:1975,Ritus:1977a,LebedevRitus:1984}.

The most general constant-field configuration is not a purely electric or magnetic background, however, but one in which both electric and magnetic fields are simultaneously present. This is the setting we are interested in and the analysis of which we broach in the present study. Beyond the potential for unique analytical behaviors during reconstruction, the configuration of parallel electric and magnetic fields investigated here represents the completely general case, in contrast to purely electric or purely magnetic backgrounds.

For one, note that the scenario of a constant and homogenous background which consists of both electric and magnetic fields is a relatively universal situation from an electrodynamics point of view. This is because for any spatially homogeneous $\vec{E}$ and $\vec{B}$ fields configuration for which the Lorentz invariant $\vec{E} \cdot \vec{B}$ is non-vanishing, an infinite number of reference frames may be found where the transformed fields are parallel to each other. Specifically, a Lorentz boost ($\vec{\beta}$) given implicitly by the relation\,\cite{LandauLifshitz1975}
\begin{equation}
\frac{\vec{\beta}}{1+\lvert\vec{\beta}\rvert^2}=\frac{\vec{E}\times \vec{B}}{\lvert \vec{E}\rvert^2+\lvert \vec{B}\rvert^2}\; ,
\end{equation}
would yield one such reference frame where the transformed fields are now parallel to each other. This is the so-called centre-of-field frame. The other inertial frames where the fields are also parallel are related to the centre-of-field frame above by arbitrary boosts along the direction in which the fields are parallel. For this reason, we will henceforth refer to constant backgrounds that consist of both electric and magnetic fields as the ``parallel fields" case. For homogeneous configurations where the fields satisfy $\vec{E}\cdot\vec{B} = 0$ but are unequal in magnitude, an inertial frame can always be found in which the background reduces to a purely electric or purely magnetic field. In this latter scenario, the relevant expressions are those of a single field and the analysis would reduce to that of previous works in the literature\,\cite{florio,dunne-harris-higher-loop,Gupta:2023gsw}. We will henceforth refer to this case as the ``perpendicular fields" configuration. In the case when $\vec{E}\cdot \vec{B} = 0$, and the fields are equal in magnitude, there are no quantum corrections. Thus, from a field configuration point of view, the case we analyse with spatially homogeneous, parallel electric and magnetic fields is completely complementary to the scenarios with purely electric or magnetic backgrounds. On the other hand, from the resummation and resurgence analysis point of view, taking the special limit where one of the fields vanishes should yield results equivalent to cases with purely electric or magnetic backgrounds. Thus, the present analysis is general and complementary.

In Sec.\,\ref{sec:review} we begin by briefly reviewing some of the relevant concepts in resurgent and resummation analyses while fixing our notations and conventions. In Sec.\,\ref{sec:2fieldEHL} we discuss the one-loop QED and SQED effective Lagrangian, and its asymptotic expansions, which will be the main starting point of our investigation. Known special-function series expressions for the real and imaginary parts are reviewed. Then, in Sec.\,\ref{sec:large-n-expansion} we discuss various theoretical aspects associated with the weak-field expansion and the Borel transform in the parallel fields case. Here, the weak-field expansion is reorganised as a single-variable asymptotic series, keeping the field ratio fixed, and the large-order behaviour of the expansion coefficients is derived using Borel dispersion relations. The parallel field Borel plane is shown to contain singularities on both real and imaginary axes, unlike the perpendicular fields case. Explicit asymptotic formulas are obtained here for QED and SQED coefficients, revealing alternating and non-alternating factorial growth. The dependence on the field ratio is shown to interpolate smoothly between the perpendicular fields magnetic and electric limits. Numerical checks confirm rapid convergence to the predicted asymptotics. Then, in Sec.\,\ref{sec:resum} we attempt to reconstruct the full parallel fields EHL from weak-field data using Pad\'{e}–Borel and Pad\'{e}–Conformal–Borel resummation techniques. The Pad\'{e}–Borel method is seen to successfully reproduce both the real and imaginary parts for moderate field ratios, thereby capturing Schwinger pair-production effects. For large field ratios, the standard Pad\'{e}–Borel resummation diverges from the special-function series expressions in the strong-field regime and fails to capture the sign-changing features in spinor QED. A conformal mapping of the Borel plane is observed to improve convergence and accuracy in QED significantly. Similar conformal mappings in the SQED EHL also lead to minor improvements in the reconstruction. Detailed numerical comparisons with the theoretical analyses and exact special-function results are presented. We summarise the main results and conclude in Sec.\,\ref{sec:conclusion}.

\section{Borel transform and resummation of asymptotic series}
\label{sec:review}

To clarify our notations and conventions, and to make the study relatively self-contained, we will briefly review some of the relevant concepts of asymptotic series, Borel transform, Borel resummation, and resurgence as applied in this study. We will review the basic theory here and, later in the relevant sections, give its extensions as used in the parallel fields case.

In perturbative treatments of quantum field theory, we often encounter solutions or expressions that are formally a series in a small expansion parameter such as the coupling constant, inverse temperature or a weak field. They generically have a form 
\begin{equation}
\mathcal{S}(z) =  \sum_{k=0}^\infty  a_k z^k\,,
\label{eq:asymseries}
\end{equation}
assuming the expansion is about $z=0$. The radius of convergence $R$ of such a formal series may be obtained as
\begin{equation}
R^{-1} = \varlimsup_{k\rightarrow\infty} |a_k|^{\frac{1}{k}} \,.
\end{equation}
Here, the limit supremum is being taken. In many cases of interest, though, the radius of convergence is zero, and then a question arises as to how even to interpret the formal series.

Though divergent, the series that appear are often asymptotic, which means that the series under an optimal truncation can approximate the function in a region of the complex plane to some accuracy. More precisely, a function $\mathbb{S}(z)$ is said to be asymptotic to the series in Eq.\,(\ref{eq:asymseries}), denoted by $\mathcal{S}(z)$
\begin{equation}
\mathbb{S}(z) \sim \mathcal{S}(z)~~~(z\rightarrow0)\;,
\end{equation}
if the difference between the two is negligible in the limit $z\rightarrow 0$ and is bounded by the last term retained in the series.
\begin{equation}
\mathbb{S}(z) - \sum_{k=0}^{N} a_k z^k \ll z^N ~~~(z \rightarrow 0)\,.
\end{equation}
Thus, optimal truncation implies that one can best approximate the function in a region of the complex plane by retaining a finite number of terms in the series, with the last term retained being the smallest. Optimal truncation can often give a good approximation to the function for some values of the small expansion parameter, but usually fails to provide a good approximation for other values of the parameter.

The starting point of resummation schemes is the question of whether one can do better and utilise all the information contained in the complete asymptotic series expansion, which the optimal truncation procedure throws away.

Towards this, the standard Borel transform of the formal series $\mathcal{S}(z)$ may be defined as
\begin{equation}
\hat{\mathcal{S}}(\zeta)\equiv \sum_{k=0}^\infty \frac{a_k}{k!}\zeta^k\,.
\end{equation}
Sometimes, a generalisation of the above, termed a general Borel transform $\widehat{\mathcal{S}}_{\lambda,N^*} (z)$, is also defined as\,\cite{jentschura_asymptotic_2000,jentschura_resummation_2001} 
\begin{equation}
    \widehat{\mathcal{S}}_{\lambda,N^*} (\zeta) = \sum_{n=0}^{N^*} \frac{a_n}{(2n+\lambda)!} \ \zeta^{2n+\lambda} \ , \ \lambda\geq 0 \; ,
    \label{eq:GBTdef}
\end{equation}
where $N^*$ denotes the order at which the Borel transform is truncated.

The asymptotic series in Eq.\,(\ref{eq:asymseries}) is said to be of Gevrey-p type if further we have
\begin{equation}
    \vert a_k \vert \leq c_0 A^{-k} (k!)^p~~\forall k\;,
\end{equation}
for some constants $c_0$ and $A$. If the series $\mathcal{S}(z)$ is of Gevrey-1 type, then by explicit construction the Borel transform $\hat{\mathcal{S}}(\zeta)$ is analytic in a domain around the origin and therefore has now a non-vanishing radius of convergence. The plane spanned by the argument ($\zeta$) of the Borel transform is referred to as the Borel plane. The singularities in the Borel plane contain much information about the true function $\mathbb{S}(z)$.

If the domain of anyticity of the Borel transform $\hat{\mathcal{S}}(\zeta)$ may be extended to include the positive real-axis, we may then define a Laplace transform integral
\begin{equation}
\text{B} [\mathcal{S}]_\gamma (z) \equiv \int_0^\infty d\zeta\, e^{-\zeta} \hat{\mathcal{S}}(z\zeta)\,.
\label{eq:bsum}
\end{equation}
$\text{B} [\mathcal{S}]_\gamma (z)$ is formally termed the Borel resummation of the asymptotic series $\mathcal{S}(z)$ denoted by Eq.\,(\ref{eq:asymseries}). Here $\gamma$ refers to the contour of integration, which in this case is the positive real axis.

In many cases, though, even when the series may in principle be Borel summable, only a finite number of coefficients in its perturbative expansion may be typically known (say, $N^*$). This severely impedes the analytic continuation of the associated Borel transform to a neighbourhood of the positive real axis. A particularly effective approach to construct accurate approximations in these cases is provided by Pad\'{e} approximants (see, for example, \cite{baker1996pade}). In a form amenable to our analysis, the relevant Pad\'{e} approximants may be defined as
\begin{equation}
\begin{split}
  &P^N_{M}\left[ \hat{S}_{\lambda,N^*} \right] \left(\zeta\right) \equiv  \frac{P_0 + P_1 \zeta + \dots + P_N \zeta^N}{1 + Q_1 \zeta + \dots Q_M \zeta^M} \\ & \sim \sum_{k=0}^{N^*} \frac{a_k}{(2k+\lambda)!} \zeta^{2k+\lambda}\equiv  \widehat{\mathcal{S}}_{\lambda,N^*} (\zeta) ~~ ( \zeta \to 0 )\; .
\end{split}
\end{equation}
Here, the $N+M+1$ coefficients in the approximant ($P_a,\,Q_b$) are determined from the $N^*+1$ coefficients $a_k$ in the finite perturbation series. The special cases with $M=N$ and $M=N+1$ are often referred to as the diagonal and off-diagonal Pad\'{e} approximants.
    
Often, the Borel transform of the series will have singularities on the real positive axis. Hence, one may define a lateral Borel sum by deforming the contour of integration to be above ($\gamma^+$) and below ($\gamma^-$) to avoid these singularities. The lateral Borel sum is then defined with these deformed contours as
\begin{equation}
\text{B} [\mathcal{S}]_{\gamma^{\pm}}(z) \equiv \int_{\gamma^{\pm}} d\zeta\, e^{-\zeta} \hat{\mathcal{S}}(z\zeta)\,.
\end{equation}
More generally, the Borel transform singularities may appear at other locations in the complex plane, say at an angle $\theta$ with respect to the positive real-axis, and the contour of integration may now be deformed above and below to avoid these singularities. This leads to the definition of a generalised lateral Borel sum
\begin{equation}
\text{B} [\mathcal{S}]_{\gamma^{\theta_\pm}}(z) \equiv \int_{\gamma^{\theta_\pm}} d\zeta\, e^{-\zeta} \hat{\mathcal{S}}(z\zeta)\,.
\end{equation}

By the dominated convergence theorem (see, for example, \cite{Bartle1995}), we may exchange the order of summation and integration as
\begin{equation}
\int_0^\infty  d\zeta\, \sum_{k=0}^\infty e^{-\zeta}\, \frac{a_k}{k!} z^k \zeta^k \stackrel{?}{=} \sum_{k=0}^\infty\, \int_0^\infty d\zeta\, e^{-\zeta} \, \frac{a_k}{k!} z^k \zeta^k
\end{equation}
only if the absolute convergence
\begin{equation}
\sum_{k=0}^\infty \, \int_0^\infty d\zeta \, \bigg\lvert e^{-\zeta}\, \frac{a_k}{k!} z^k \zeta^k \bigg\rvert < \infty \,,
\end{equation}
is satisfied. In many problems, this latter condition, required by the dominated convergence theorem, is not satisfied. Hence, in general, it is not true that the Borel resummation $\text{B} \mathcal{S}_\gamma (z)$ of the asymptotic series is identical to the latter
\begin{equation}
\text{B} [\mathcal{S}]_\gamma (z) \neq \mathcal{S}(z) \; .
\end{equation}
Nevertheless, what is guaranteed by construction is that the asymptotic expansion of the Borel resummation $  B[\mathcal {S}]_\gamma (z)$ identically matches the asymptotic series $\mathcal{S}(z)$ of $\mathbb{S}(z)$.

Given that disparate functions may admit the same asymptotic series expansions, if they differ by sub-dominant terms (see, for instance, \cite{BenderOrszag1999}), it is interesting to ask when one may expect to have the Borel resummation fully reconstructing the true function $\mathbb{S}(z)$
\begin{equation}
\text{B}[ \mathcal{S}]_\gamma (z) \stackrel{?}{=} \mathbb{S}(z) \; .
\end{equation}
This is especially pertinent since the asymptotic expansion of the Borel resummation in Eq.\,(\ref{eq:bsum}) matches the asymptotic series in Eq.\,(\ref{eq:asymseries}) of the true function $\mathbb{S}(z)$.

The criterion for this is given by the Nevanlinna-Sokal-Watson theorem\,\cite{Nevanlinna1919, Sokal1980}. It states that if $\mathbb{S}(z)$ is an analytic function in a open disc of radius $R$ whose center is at $z=R$ and has an asymptotic expansion in that region which of Gevrey-1 type (i.e. $a_k \sim A^{-k} k!$ for large $k$), then the Borel transform $\hat{\mathcal{S}}(\zeta)$ converges for $|\zeta| < A $ and has an analytic continuation which is bounded as
\begin{equation}
    \hat{\mathcal{S}}(\zeta) \leq c_1 e^{\frac{|\zeta|}{R}} \,,
\end{equation}
for some constant $c_1$, in a band about the positive real axis. Furthermore, one then has
\begin{equation}
\text{B}[\mathcal{S}]_\gamma(z) = \mathbb{S}(z)
\end{equation}
in the open disc of radius R.

One of the impediments to applying the above theorem is that in many physical scenarios, one does not know the analyticity properties of the master function $\mathbb{S}(z)$ that we are attempting to reconstruct. In the present context, for general quantum field theories, it is therefore typically unknown whether an observable or quantity of interest can be reconstructed faithfully from its small-coupling perturbative expansion or weak-field expansions\,\cite{Lipatov1977}.

As mentioned earlier, there currently exist studies applying techniques drawn from resurgence and Borel resummation to perpendicular fields scenarios in SQED and QED\,\cite{Dunne:1999uy, florio,dunne-harris-higher-loop,DunneHarrisInhom,Gupta:2023gsw}. Our focus in the present work, in similar contexts, is on the general case with constant electric and magnetic fields simultaneously present. In brief, our primary strategy to tackle the parallel fields scalar and spinor QED cases, as we shall expound in more detail, will be to rewrite the weak-field expansions in the form
\begin{equation}
    \mathcal{S}(\alpha,\beta)=  \sum_{m=0}^\infty \sum_{n=0}^\infty  c_m \, d_n \,\alpha^m \beta^n \,,
\end{equation}
as
\begin{equation}
    \mathcal{S}(\alpha,\kappa)=  \sum_{m=0}^\infty   a_m \, P_m(\kappa) \,\alpha^m \,,
\end{equation}
with $\mathcal{P}_m(\kappa)$ taking the form of a fixed-order polynomial in $\kappa=\beta/\alpha$. We will then treat this as an asymptotic series in $\alpha$, keeping $\kappa$ fixed. This approach is relatively general in that parameters $\eta$ and $\kappa$ can be found for arbitary configurations of constant and homogenous electromagnetic background. Some resurgent information of the theory is expected to be lost by treating the field strength ratio $\kappa$ as a constant, but nevertheless reconstruction techniques like Borel resummations reveal non-trivial non-perturbative properties of the theory, as we'll see in Secs.\,\ref{sec:large-n-expansion} and \ref{sec:resum}.

\section{The QED and SQED Euler-Heisenberg Lagrangian for parallel fields}
\label{sec:2fieldEHL}

The EHL in QED and SQED with pure electric and magnetic field backgrounds have been extensively studied, in the context of resurgence, in closed form\,\cite{florio,Dunne:1999uy} and at higher loop orders\,\cite{dunne-harris-higher-loop,Gupta:2023gsw}. In contrast, to the best of our knowledge, resurgent aspects of the Euler–Heisenberg Lagrangian in constant electromagnetic backgrounds featuring simultaneous electric and magnetic fields remain unexplored in the existing literature.

Recently, an intriguing dispersive integral representation of the one-loop Heisenberg–Euler effective Lagrangian in a constant background with both an electric and magnetic field has been found\,\cite{Dunne:2025cyo}. It explicitly reveals the relationship between the real and imaginary parts, showing that the imaginary part governing Schwinger pair production is naturally expressed in terms of the quantum dilogarithm, while the real part appears as a dispersion integral involving both the quantum dilogarithm and its modular dual. This structure provides a concrete realisation of some earlier ideas by Lebedev and Ritus on dispersion representations of the effective Lagrangian in a generic constant background\,\cite{LebedevRitus1978}. General scenarios involving simultaneously both electric and magnetic background fields in QED and SQED, therefore, exhibit a wealth of intriguing theoretical features that remain only partially understood and continue to merit detailed investigation. It is also in this broad context that our investigation involving parallel electric and magnetic fields, describing the most general scenario with a constant electromagnetic background, complements the perpendicular fields resurgent and resummation analyses existing in the literature\,\cite{florio,Dunne:1999uy,dunne-harris-higher-loop,Gupta:2023gsw}. 

The effective one-loop Lagrangian for quantum electrodynamics was first given by Heisenberg and Euler in terms of the electrodynamic invariants $\mathcal{F}$ and $\mathcal{G}$ as (in natural units $c = \hbar = 1$)\,\cite{Heisenberg-Euler-OG-Paper,Schwinger-Paper}
\begin{multline}
    \mathcal{L}^{(1)}_{\text{QED}} = - \frac{1}{8 \pi^2} \int_0^{\infty} \frac{dt}{t^3} \ e^{- m^2 t} \times \\\left[q^2 t^2 \mathcal{G} \ \frac{\text{Re}\ \cosh\left(q t \sqrt{2 \mathcal{F}+2i \mathcal{G}} \right)}{\text{Im}\ \cosh\left(q t \sqrt{2 \mathcal{F}+2i \mathcal{G}} \right)} -1 - \frac{2}{3} q^2t^2 \mathcal{F}  \right] \; ,
    \label{eq:general_two-field-ehl-qed}
\end{multline}
where $\mathcal{F}$ and $\mathcal{G}$ are given by
\begin{subequations}
    \begin{align}
        &\mathcal{F} = \frac{1}{4} F_{\mu \nu} F^{\mu \nu} = \frac{\mathbf{B}^2 - \mathbf{E}^2}{2} \; , \\
        &\mathcal{G} = \frac{1}{4} F_{\mu \nu} \Tilde{F}^{\mu \nu} = - \ \mathbf{E} \cdot \mathbf{B} \; .
    \end{align}
\end{subequations}
$F^{\mu \nu}$ and $\Tilde{F}^{\mu \nu}$ are the electromagnetic and dual electromagnetic tensors respectively. Eq.\,(\ref{eq:general_two-field-ehl-qed}) can also be expressed in terms of the parallel frame magnetic and electric fields $\eta$ and $\epsilon$ as\,\cite{dunne-book}
\begin{multline}\label{eq:two-field-ehl-qed}
    \mathcal{L}^{(1)}_{\text{QED}}(\epsilon,\eta) = -\frac{1}{8\pi^2} \int_0^{\infty} \frac{dt}{t^3} \ e^{-m^2 t} \times \\ \left[ \frac{q^2t^2 \eta \epsilon}{\tanh(q \eta t) \tan(q \epsilon t)}  -1 - \frac{q^2 t^2}{3} (\eta^2 - \epsilon^2) \right] \; ,
\end{multline}
where $\eta$ and $\epsilon$ are given by
\begin{subequations}\label{eq:eta_epsilon_def}
    \begin{align}
        & \eta = \sqrt{\sqrt{\mathcal{F}^2 + \mathcal{G}^2}+\mathcal{F}} \; , \\
        & \epsilon = \sqrt{\sqrt{\mathcal{F}^2 + \mathcal{G}^2}- \mathcal{F}} \; .
    \end{align}
\end{subequations}
The substitution of the limits $\epsilon \to 0$ ($\eta \to 0$) in Eq.\,(\ref{eq:two-field-ehl-qed}) reduces $\mathcal{L}^{(1)}_{\text{QED}}$ to the EHL of a pure magnetic (electric) field background. 

Although the integral form in Eq.\,(\ref{eq:two-field-ehl-qed}) does not have a known closed form, a series representation for the real and imaginary parts of the integral is known\,\cite{Mielniczuk_1982,Valluri-special-functions,pair-production-rate,convergent-expansion-of-ehl-1,convergent-expansion-of-ehl-2} and is given by 
\begin{equation}\label{eq:ehl-numerical-real-part-qed}
    \text{Re} \ \mathcal{L}^{(1)}_{\text{QED}} = - \frac{q^2 \epsilon \eta}{4 \pi^3} \sum_{k=1}^{\infty}  \Big(\mathcal{T}_k^{\text{sp}} + \Upsilon_k^{\text{sp}} \Big) \; ,
\end{equation}
where $\mathcal{T}_k^{\text{sp}}$ and $\Upsilon_k^{\text{sp}}$ are given by
\begin{subequations}
    \begin{align}
        \mathcal{T}^{\text{sp}}_k &= \frac{1}{k} \coth \left( \frac{k \pi \epsilon}{\eta} \right) \left[\text{Ci} \left(\frac{k \pi m^2}{q \eta} \right) \cos \left(\frac{k \pi m^2}{q \eta} \right) \right. \nonumber \\ &\left. + \text{Si}\left(\frac{k \pi m^2}{q \eta} \right) \sin \left(\frac{k \pi m^2}{q \eta} \right)  \right] \; ,  \\
        \Upsilon^{\text{sp}}_k &= - \frac{1}{2k} \coth \left(\frac{k \pi \eta}{\epsilon} \right) \left[\text{Ei} \left(-\frac{k \pi m^2}{q \epsilon} \right) \exp \left(\frac{k \pi m^2}{q \epsilon} \right) \right. \nonumber \\ &\left. + \text{Ei} \left(\frac{k \pi m^2}{q \epsilon} \right) \exp \left(-\frac{k \pi m^2}{q \epsilon} \right)\right] \; ,
    \end{align}
\end{subequations}
and
\begin{equation}\label{eq:ehl-imaginary-instanton-expansion-qed}
    \text{Im} \ \mathcal{L}^{(1)}_{\text{QED}} = \frac{q^2 \epsilon \eta }{8 \pi^2} \sum_{k=1}^{\infty} \frac{1}{k} \coth\left(\frac{k \pi \eta}{\epsilon}\right) \exp \left(- \frac{k \pi m^2}{q \epsilon} \right)  \;.
\end{equation}
The unique form of the imaginary part of the QED Lagrangian with a hyperbolic cotangent in the prefactor\,\cite{Nikishov1969, BunkinTugov1970} may be motivated, for instance, by a worldline instanton derivation\,\cite{Korwar:2018euc}. 

The real part of $\mathcal{L}^{(1)}_{\text{QED}}$ has been defined in terms of the special functions $\text{Ci}(x)$, $\text{Si}(x)$, and $\text{Ei}(x)$ which are the Cos, Sin and Exponential integrals and are given by\,\cite{korn2013mathematical}
\begin{subequations}
    \begin{align}
        &\text{Ci}(x) = - \int_x^{\infty} dt \ \frac{\cos(t)}{t} \; ,\\
        &\text{Si}(x) = - \int_x^{\infty} dt \ \frac{\sin(t)}{t} \; , \\
        &\text{Ei}(x) = - \int_{-x}^{\infty} dt \ \frac{\exp(-t)}{t} \; .
    \end{align}
\end{subequations}

Like the spinor case, the one-loop scalar QED EHL for parallel fields can also be written in integral form\,\cite{Weisskopf1936} as
\begin{multline}\label{eq:ehl-integral-scalar}
    \mathcal{L}_{\text{SQED}}^{(1)}(\epsilon,\eta) = \frac{1}{16\pi^2} \int_0^{\infty} \frac{dt}{t^3} e^{-m^2 t} \times \\ \left[\frac{q^2t^2\eta \epsilon}{\sinh (q\eta t) \sin(q\epsilon t)} - 1 +  \frac{q^2 t^2}{6} (\eta^2 - \epsilon^2) \right] \; .
\end{multline}
In terms of an infinite sum of special functions, the real and imaginary parts of the scalar QED EHL are given by\,\cite{scalar-special-functions} 
\begin{equation}\label{eq:ehl-numerical-real-part-sqed}
    \text{Re}\ \mathcal{L}_{\text{SQED}}^{(1)} = - \frac{q^2 \epsilon \eta}{8 \pi^3} \sum_{k=1}^{\infty} \Big( \mathcal{T}_k^{\text{sc}} + \Upsilon_k^{\text{sc}} \Big)   \; ,
\end{equation}
where
\begin{subequations}
    \begin{align}
        \mathcal{T}_k^{\text{sc}} &= \frac{(-1)^{k-1}}{k} \frac{1}{\sinh \left(k\pi \epsilon/\eta \right)} \times \nonumber \\ &\left[\text{Ci} \left(\frac{k \pi m^2}{q \eta} \right) \cos \left(\frac{k \pi m^2}{q \eta} \right) \right. \nonumber \\ &\left. + \text{Si}\left(\frac{k \pi m^2}{q \eta} \right) \sin \left(\frac{k \pi m^2}{q \eta} \right)  \right] \; ,\\
        \Upsilon_k^{\text{sc}} &= - \frac{(-1)^{k-1}}{2k} \frac{1}{\sinh(k\pi\eta/\epsilon)} \times \nonumber \\&\left[\text{Ei} \left(-\frac{k \pi m^2}{q \epsilon} \right) \exp \left(\frac{k \pi m^2}{q \epsilon} \right) \right. \nonumber \\ & \left. + \text{Ei} \left(\frac{k \pi m^2}{q \epsilon} \right) \exp \left(-\frac{k \pi m^2}{q \epsilon} \right)\right] \; ,
    \end{align}
\end{subequations}
and
\begin{multline}\label{eq:ehl-imaginary-instanton-expansion-sqed}
        \text{Im} \ \mathcal{L}_{\text{SQED}}^{(1)} = \frac{q^2 \epsilon \eta}{16\pi^2} \sum_{k=1}^{\infty} \frac{(-1)^{k-1}}{k} \frac{1}{\sinh ( k \pi \eta/\epsilon)} \times \\  \exp \left(- \frac{k \pi m^2}{q \epsilon} \right) \; .
\end{multline}
The imaginary part of the scalar QED Lagrangian again has a characteristic prefactor \,\cite{Popov1972a,Popov1972b}--now with a hyperbolic sine inverse instead of a hyperbolic cotangent. The functional form may again be seen most directly from a worldline instanton analysis of scalar QED\,\cite{Korwar:2018euc}.

The imaginary parts of $\mathcal{L}^{(1)}_{\text{QED}}$ and $\mathcal{L}^{(1)}_{\text{SQED}}$ in Eqs.\,(\ref{eq:ehl-imaginary-instanton-expansion-qed}) and (\ref{eq:ehl-imaginary-instanton-expansion-sqed}) indicate the possibility of the decay of the vacuum and the spontaneous creation of particle-antiparticle pairs, a phenomenon known as Schwinger pair production. This nonperturbative imaginary contribution to $\mathcal{L}^{(1)}$ cannot be probed by conventional perturbative expansions, and therefore resurgent analysis is essential to examine the true functional nature of the EHL. However, resurgent analysis in these contexts would seem to be complicated by the presence of the two variables $\eta$ and $\epsilon$, whereas a resurgent analysis is usually only done for a resurgent function of a single variable.

In the following sections, we will extrapolate the conventional single-variable resurgent analysis to include two variables. Furthermore, we will apply Pad\'{e}-Borel and Pad\'{e}-Conformal-Borel reconstructions of the parallel fields EHL to find a functional approximation to $\mathcal{L}^{(1)}_{\text{QED}}$ and $\mathcal{L}^{(1)}_{\text{SQED}}$. We will compare these to the special function series representations given in Eqs.\,(\ref{eq:ehl-numerical-real-part-qed}), (\ref{eq:ehl-imaginary-instanton-expansion-qed}), (\ref{eq:ehl-numerical-real-part-sqed}), and (\ref{eq:ehl-imaginary-instanton-expansion-sqed}) in order to determine the accuracy of these approximations. For simplicity, henceforth, we will set the mass $m$ and the electric charge $q$ to 1, i.e., $q=m=1$.

\section{Large-N expansion of weak-field coefficients in the QED and SQED parallel fields scenarios}\label{sec:large-n-expansion}

The programme of resurgence relates the apparent divergence of the weak-field expansions of the spinor and scalar EHLs to the nonperturbative phenomenon of Schwinger pair production, the rate of which is encoded in the imaginary part of the EHL. For the perpendicular fields EHL, this connection has been firmly established in both spinor and scalar QED\,\cite{dunne-harris-higher-loop,Gupta:2023gsw} through the use of Borel dispersion relations. These relations provide a systematic prescription for extracting the $k^{\text{th}}$ instanton contribution to the Schwinger pair-production rate from the residue at the $k^{\text{th}}$ pole of the Borel transform of the divergent weak-field series, which in turn can be extracted from the asymptotic expansion of the weak-field coefficients $a_N$ as $N\to \infty$, which we will refer to as ``large-N asymptotics".

In the parallel fields, one-loop EHL, the instanton expansions governing the Schwinger pair-production rate in scalar and spinor QED are known to take the forms given in Eqs.\,(\ref{eq:ehl-imaginary-instanton-expansion-qed}) and (\ref{eq:ehl-imaginary-instanton-expansion-sqed}). This naturally raises the question of whether resurgence techniques can be generalised to the case of multiple variables and applied to the weak-field expansion of the parallel fields EHL. In this section, we show that the seemingly divergent weak-field expansions of the parallel fields EHLs likewise encode detailed information about the nonperturbative imaginary parts associated with parallel fields spinor and scalar Schwinger pair production.

Let us first examine the weak-field expansion of the integral expression of $\mathcal{L}^{(1)}_{\text{QED}}$ as given in Eq.\,(\ref{eq:two-field-ehl-qed}) in the limit $\epsilon, \eta \to 0$. The resultant formal series has terms of the form $\eta^n \epsilon^m$, where $n,m>0$ are integers. We will simplify the problem by introducing the variable $\kappa = \epsilon/\eta$ and rewriting each term of the form $\eta^n \epsilon^m$ as $\eta^{n+m} \kappa^m$. Gathering terms of order $\eta^n$ will give us a weak field expansion of the form
\begin{equation}\label{eq:weak-field-expansion-spinor}
    \mathcal{L}^{(1)}_{\text{QED}}(\eta,\kappa) \sim - \frac{\eta^4}{8 \pi^2} \ \sum_{n=0}^{\infty} a_n^{\text{sp},(1)} \ P_n^{\text{sp}}(\kappa) \ \eta^{2n} \ , \ \eta \to 0 \;, 
\end{equation}
where $P_n(\kappa)$ is a polynomial in $\kappa$ of degree $2n+4$. From now on, we will treat $\kappa$ as a constant (notice that this requires the limit of $\epsilon \to 0$ to be taken such that the ratio of $\epsilon$ and $\eta$ remains constant). We will also drop the superscript $(1)$, indicating that the weak-field coefficients belong to the one-loop EHL, because we will only be dealing with one-loop EHLs in this work.

\begin{figure*}[!ht]
    \centering
    \begin{subfigure}{0.45\linewidth}
        \centering
        \includegraphics[width=\linewidth]{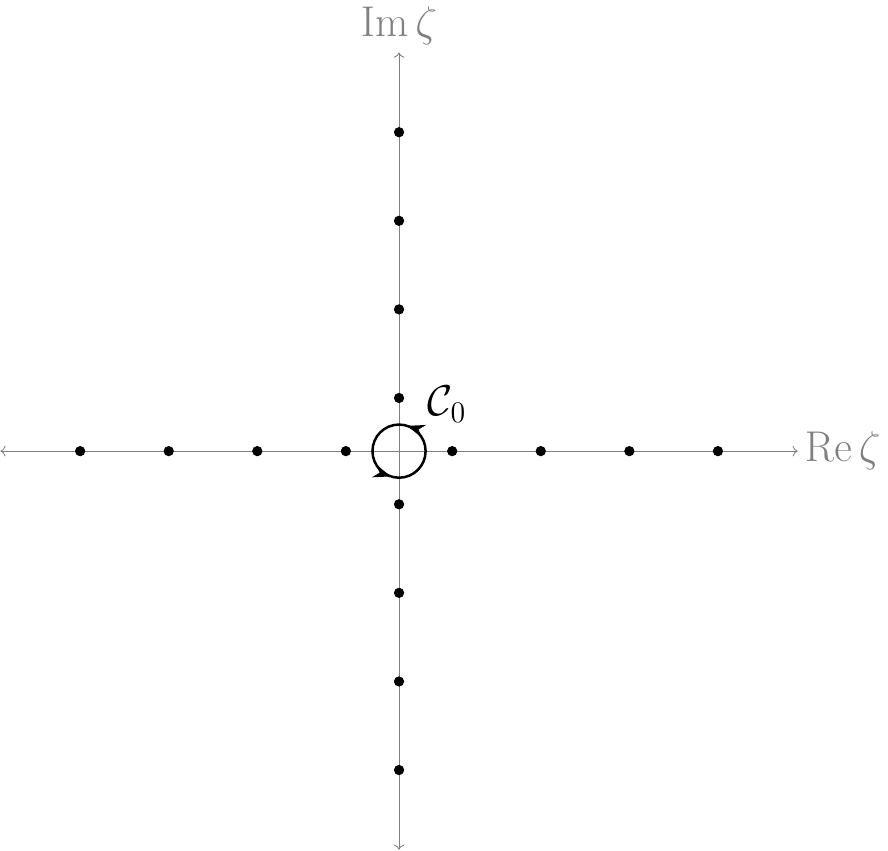}
        \caption{}
        \label{fig:contour_plot_1}
    \end{subfigure}
    \begin{subfigure}{0.45\linewidth}
        \centering
        \includegraphics[width=\linewidth]{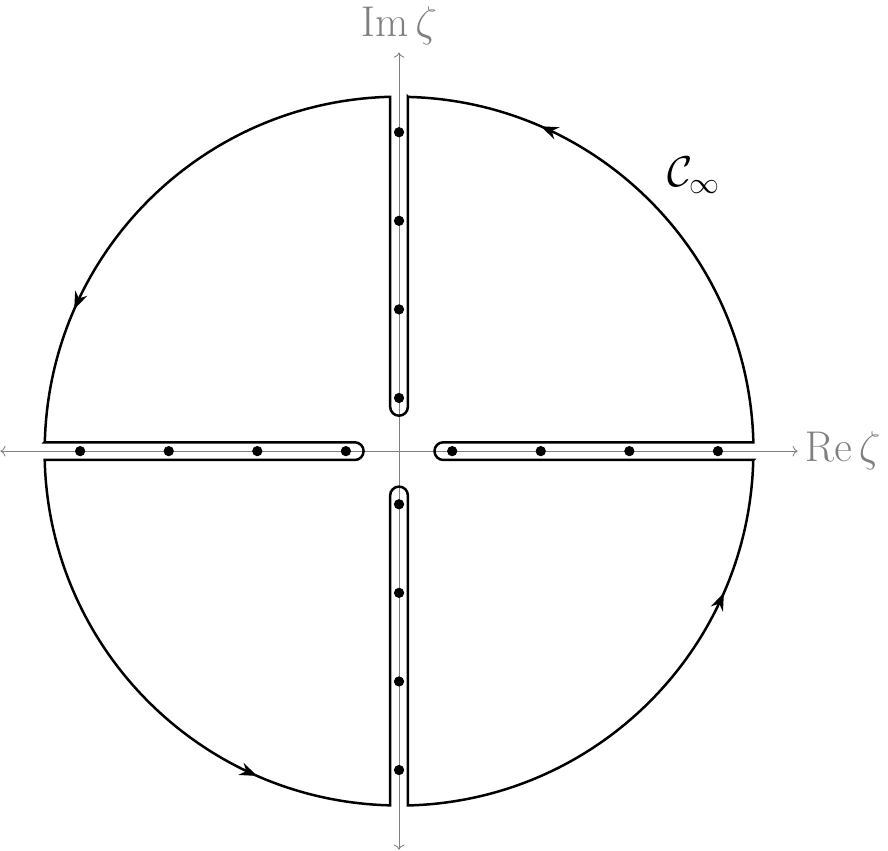}
        \caption{}
        \label{fig:contour_plot_2}
    \end{subfigure}
    \caption{Contour integrals for Borel dispersion relations for the parallel fields scalar and spinor EHLs. The singularities on the Borel plane occur at $k\pi i$ and $k\pi/\kappa$ with $k \in \mathbb{Z}-\{0\}$. Notice that, unlike the perpendicular fields case, the Borel plane has singularities both on the real and imaginary axes, which leads to a contour $\mathcal{C}_{\infty}$ that has slots both on the real and imaginary axes instead of just the imaginary axis, as is the case in the perpendicular fields Borel plane.}
    \label{fig:contout_plots}
\end{figure*}

To derive the asymptotic behaviour of the coefficients $a_n^{\text{sp}}$ and $P_n^{\text{sp}}$ as $n\to \infty$ (i.e., the large-N asymptotics), we will use the well-known techniques that have previously been used to derive similar asymptotic limits for many different weak-field expansions (see, for eg.,\,\cite{Marino_2015,Gupta:2023gsw}). Consider the formal series
\begin{equation}\label{eq:borel-sum-spinor}
    \mathcal{S}^{\text{sp}}(\eta,\kappa) = \sum_{n=0}^{\infty} a_n^{\text{sp}} \ P_n^{\text{sp}}(\kappa) \ \eta^{2n} \; .
\end{equation}
For this formal series, define the Borel transform
\begin{equation}\label{eq:borel-transform-def}
    \widehat{\mathcal{S}}^{\text{sp}}_{\lambda} (\zeta, \kappa) = \sum_{n=0}^{\infty} \frac{a_n^{\text{sp}} P_n(\kappa)}{(2n+\lambda)!} \ \zeta^{2n+\lambda} \; .
\end{equation}
The parameter $\lambda$ can be varied to obtain optimal convergence \,\cite{Gupta:2023gsw}. This has also been called the ``asymptotically improved" Borel sum in the literature\,\cite{jentschura_asymptotic_2000,jentschura_resummation_2001}. Analysing the integrand in Eq.\,(\ref{eq:two-field-ehl-qed}) with the variable change $t \to t/ q\eta$ tells us that the integrand has a simple pole at $t= k \pi i$ and $t=k \pi/\kappa$ for all integer values of $k$ except $k=0$. As a consequence, we also expect the Borel sum $\widehat{S}^{\text{sp}}$ to also have simple poles at these locations in the complex $\zeta$ plane, as demonstrated in Fig.\,(\ref{fig:contout_plots}).

Additionally, notice that the integrand in Eq.\,(\ref{eq:two-field-ehl-qed}) has a symmetry under the transformation $\eta \to - i \epsilon$, $\epsilon \to -i \eta$ (this will be a generic property of the EHLs at all orders because the EHLs are functions of the gauge invariant variables $\mathcal{F}$ and $\mathcal{G}^2$, which are invariant under this symmetry). Because of this symmetry, the polynomial $P_n^{\text{sp}}(\kappa) = p_0 + p_1 \kappa^2 + \dots + p_{n+2} \kappa^{2n+4}$ will have the corresponding symmetry property 
\begin{equation}
    p_r = (-1)^{n+2} \  p_{n+2-r} \; .
\end{equation}
Using this property, we see that the Borel transform $\widehat{\mathcal{S}}_{\lambda}^{\text{sp}}$ should have the symmetry property
\begin{equation}\label{eq:symmetry-of-borel-transform}
    \widehat{\mathcal{S}}_{\lambda}^{\text{sp}} \left(-i \kappa \zeta, \ \frac{1}{\kappa} \right) = (-i \kappa)^{\lambda-4} \ \widehat{\mathcal{S}}_{\lambda}^{\text{sp}} (\zeta, \kappa) \;.
\end{equation}
Now, let us assume that the singularity structure of the Borel transform $\widehat{\mathcal{S}}^{\text{sp}}_{\lambda}$ around the singularity $k \pi/\kappa$ on the real axis is given by
\begin{equation}\label{eq:borel-transform-singularity-structure-1}
    \widehat{\mathcal{S}}^{\text{sp}}_{\lambda} \left(\pm\frac{k \pi}{\kappa} + \xi , \ \kappa \right) = (\pm1)^{\lambda-1} \ \frac{C_k}{\xi} \ \mathcal{F}_k(\kappa) \; .
\end{equation}
where the constant $C_k$ and the function $\mathcal{F}_k(\kappa)$ are to be determined. Using the symmetry of the Borel transform in Eq.\,(\ref{eq:symmetry-of-borel-transform}), we can infer the singularity structure of a pole on the imaginary axis at $i k \pi$ to be
\begin{equation}\label{eq:borel-transform-singularity-structure-2}
    \widehat{\mathcal{S}}^{\text{sp}}_\lambda (\mp i k \pi + \xi , \kappa) = (\pm1)^{\lambda-1} \left( -\frac{i}{\kappa} \right)^{\lambda-3} \ \frac{C_k}{\xi} \ \mathcal{F}_k \left(\frac{1}{\kappa} \right) \; .
\end{equation}
Now, we perform a contour integral around the contour $\mathcal{C}_0$ as is depicted in Fig.\,(\ref{fig:contour_plot_1}) in the following manner:-
\begin{equation}
    \frac{a_n^{\text{sp}} P_n(\kappa)}{\Gamma(2n+\lambda+1)} = \frac{1}{2\pi i} \oint_{\mathcal{C}_0} \frac{\hat{\mathcal{S}}_\lambda^{\text{sp}}(\zeta)}{\zeta^{2n+\lambda+1}} d \zeta \; .
\end{equation}
Now, we can deform the contour $\mathcal{C}_0$ in Fig.\,(\ref{fig:contour_plot_1}) to the Hankel-like contour $\mathcal{C}_\infty$ in Fig.\,(\ref{fig:contour_plot_2}) such that all the poles on the real and imaginary axes are avoided. The only contribution to $\mathcal{C}_\infty$ comes from the slots in the contour on the real and imaginary axes, and these can be calculated by summing up the residues of all the singularities on the real and imaginary axes. The singularity structure of the Borel transforms in Eqs.\,(\ref{eq:borel-transform-singularity-structure-1}) and (\ref{eq:borel-transform-singularity-structure-2}) will lead to the relation
\begin{equation}\label{eq:large_n_behaviour_bdr}
\begin{split}
    &\frac{a_n^{\text{sp}} P_n(\kappa)}{\Gamma(2n+\lambda+1)} = -2 \sum_{k=1}^{\infty} \frac{C_{k}}{(k\pi/\kappa)^{2n+\lambda+1}}\Biggl ( \mathcal{F}_{k}(\kappa) \\ &   + (-1)^{\lambda-1}(i\kappa)^{-(2n+\lambda+1)} \left(-\frac{i}{\kappa} \right)^{\lambda-3} \mathcal{F}_{k} \left(\frac{1}{\kappa} \right) \Biggr )    \;, \\
    & \qquad = -2 \sum_{k=1}^{\infty} \frac{C_k}{(k\pi/\kappa)^{2n+\lambda+1}} \Biggl ( \mathcal{F}_k(\kappa) \\ &+ (-1)^n \kappa^{-2(n+\lambda-1)} \mathcal{F}  \left( \frac{1}{\kappa} \right) \Biggr )
\end{split}
\end{equation} 

We can also deduce the form of the function $\mathcal{F}_k$ and the constant $C_k$ by using the singularity structure in Eq.\,(\ref{eq:symmetry-of-borel-transform}) by computing the imaginary part of the Borel sum, which is given by
\begin{equation}
    \text{B}_{\lambda} [\mathcal{S}](\eta,\kappa) = \eta^{-\lambda-1} \int_0^{\infty (1+i \epsilon)} dt \ e^{-t/\eta} \ \ \widehat{\mathcal{S}}_\lambda (\eta,\kappa)
\end{equation}
Doing the computation gives us
\begin{equation}\label{eq:imaginary_part_from_bdr}
    \text{Im} \ B \mathcal{S}^{\text{sp}} (\eta,\kappa) = -\frac{\pi }{\eta^2} \sum_{k=1}^{\infty} \exp \left(- \frac{k \pi}{\kappa \eta} \right) C_{k} \mathcal{F}_{k}(\kappa) \; .
\end{equation}
It is apparent from the structure of Eqs.\,(\ref{eq:large_n_behaviour_bdr}) and (\ref{eq:imaginary_part_from_bdr}) that the prefactors $C_k$ and $\mathcal{F}_k(\kappa)$ appearing in the instanton expansion of the imaginary part are encoded in the large-N behaviour of the weak-field coefficients. In particular, the prefactors associated with the $k^{\text{th}}$ instanton contribution are implicitly contained in the coefficient multiplying the $(k\pi)^{-2n}$ term in the large-N expansion. This fact suggests that the parallel fields EHLs are suitable for resurgent analysis.

For the case of the one-loop EHL for QED, the exact form of the instanton expansion is known as given in Eq.\,(\ref{eq:ehl-imaginary-instanton-expansion-qed}), and so we can compute the parallel fields large-N expansion for this case. Comparing Eq.\,(\ref{eq:imaginary_part_from_bdr}) to the imaginary part of the QED EHL in Eq.\,(\ref{eq:ehl-imaginary-instanton-expansion-qed}), we get that
\begin{subequations}\label{eq:temp1508241}
    \begin{align}
        & C_k = \frac{1}{\pi k } \; , \\
        & \mathcal{F}_k (\kappa) = \kappa \coth \left(\frac{k \pi}{\kappa} \right) \; .
    \end{align}
\end{subequations}
Substituting Eq.\,(\ref{eq:temp1508241}) and $\lambda =1$ (this choice is made because the weak field coefficients for the one-loop EHLs grow as $\sim \Gamma(2n+2)$\,\cite{dunne-harris-higher-loop,Gupta:2023gsw}) in Eq.\,(\ref{eq:large_n_behaviour_bdr}), we have the final expression
\begin{multline}\label{eq:large-N-behaviour-spinor}
    \frac{a_n^{\text{sp}} P_n(\kappa)}{\Gamma(2n+2)} = - 2\kappa^2 \sum_{k=1}^\infty \frac{1}{(k\pi)^{2n+3}}\left[\kappa^{2n+1} \coth \left(\frac{k \pi}{\kappa} \right) \right. \\ \left. + \frac{(-1)^n}{\kappa} \coth( k \pi \kappa) \right]
\end{multline}

From the structure of Eq.\,($\ref{eq:large-N-behaviour-spinor}$) it is apparent that the instanton prefactor, $\coth(k\pi\epsilon/\eta)$, which appears in the imaginary part of the spinor-QED EHL is mirrored in the coefficients of exponentially small ``instanton" contributions of the form $(k\pi)^{-2n}$ in the large-$N$ expansion of the weak-field coefficients of the EHL. This observation supports the resurgence paradigm, whereby nonperturbative contributions are encoded within the seemingly divergent weak-field series expansion of the EHL. Specifically, the large-N structure of the weak field coefficients suggests a two-variable generalisation of a Gevrey-1 type series, making it amenable to Pad\'{e}-Borel reconstruction. In the following section, we detail the application of Pad\'{e}-Borel and Pad\'{e}-Conformal-Borel resummation techniques to extract the nonperturbative behaviour that is seemingly hidden within the weak-field coefficients of the EHLs.

There are a few interesting features of the asymptotic limit found in Eq.\,(\ref{eq:large-N-behaviour-spinor}). The summand in this equation has two parts: one which has the alternating behaviour of $(-1)^n$ and the other which is non-alternating. For $\kappa<1$, we find that the alternating part of the summand dominates and therefore the weak-field expansion for small values of $\kappa$ is predominantly alternating. As $\kappa \to 0$, we get that
\begin{equation}
\begin{split}
    \lim_{\kappa \to 0} \frac{a_n^{\text{sp}}P_n(\kappa)}{\Gamma(2n+2)} &= - 2\kappa^2 \sum_{k=1}^{\infty}  \frac{1}{(k\pi)^{2n+3}} \frac{(-1)^n}{\kappa} \cdot \frac{1}{k \pi \kappa} \\
    &= -2\sum_{k=1}^{\infty}  \frac{(-1)^n}{(k\pi)^{2n+4}}
\end{split}
\end{equation}
This matches the well-known large-N expansion for the weak field coefficients of the QED EHL in a purely magnetic background\,\cite{dunne-harris-higher-loop}. Since the Borel transform of the weak-field expansion in a purely magnetic background is alternating and therefore convergent, it doesn't have singularities along the real axis. Consequently, the Euler-Heisenberg Lagrangian (EHL) in this regime does not develop an imaginary part. This result is consistent with the physical requirement of an electric field component to facilitate the Schwinger mechanism for pair production. When $\kappa>1$, the non-alternating part of the summand dominates, and thus the weak-field expansion is predominantly non-alternating. When $\kappa \to \infty$, 
\begin{equation}
\begin{split}
    \lim_{\kappa \to \infty} \frac{a_n^{\text{sp}}P_n(\kappa)}{\Gamma(2n+2)} &= - 2\kappa^2 \sum_{k=1}^{\infty}  \frac{\kappa^{2n+1}}{(k\pi)^{2n+3}} \cdot \frac{\kappa}{k\pi} \\
    &= -2 \kappa^{2n+4}\sum_{k=1}^{\infty}  \frac{1}{(k\pi)^{2n+4}} \; ,
\end{split}
\end{equation}
which matches the large-N expansion for the weak-field coefficients in a purely electric background. In this limit, the weak-field expansion becomes completely non-alternating. When $\kappa=1$, the alternating and the non-alternating terms of the summand cancel exactly when $n$ is odd, and therefore we end up with a weak-field expansion in which only the coefficients of terms of the form $\eta^{4n}$ are non-zero. This is because when $\kappa=1$ or equivalently, $\mathcal{F} = 0$, we have that $\eta^2  = |\mathcal{G}|$ from Eq.(\ref{eq:eta_epsilon_def}). Since the EHL only has parity-conserving terms, the expansion of the EHL only has even powers of $\eta^2$.

\begin{figure*}[!ht]
    \centering
    \begin{subfigure}{0.48\textwidth}
        \includegraphics[width=\textwidth]{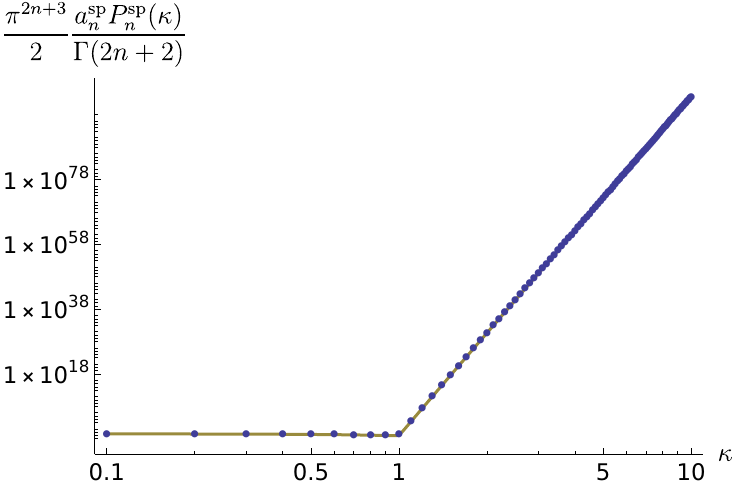}
        \caption{}
        \label{fig:large-n-behaviour-spinor}
    \end{subfigure}
    \begin{subfigure}{0.48\textwidth}
        \includegraphics[width=\textwidth]{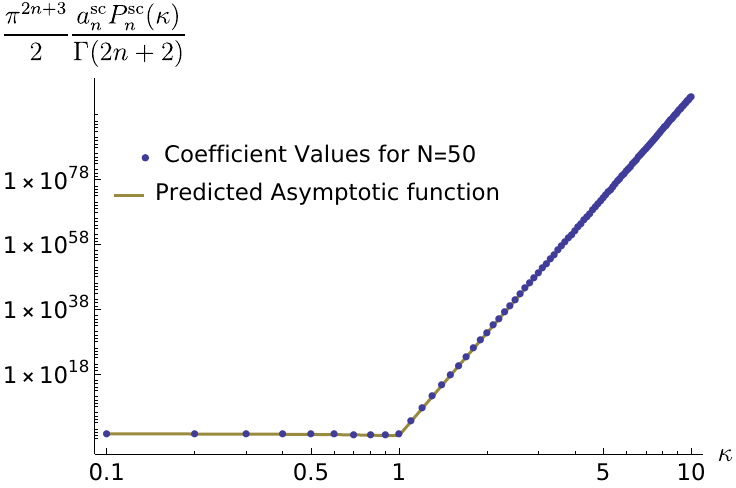}
        \caption{}
        \label{fig:large-n-behaviour-scalar}
    \end{subfigure}
    \caption{Values of the weak-field expansion coefficient for $n=50$ plotted as a function of $\kappa$ for the (a) spinor, and (b) scalar QED parallel fields EHL compared with the expected large-N expansion computed in Eqs.\,(\ref{eq:large-N-behaviour-spinor}) and (\ref{eq:large-N-behaviour-scalar}). Note that these are plots of absolute values.}
    \label{fig:large-n-behaviour}
\end{figure*}

\begin{figure*}[!ht]
    \centering
    \begin{subfigure}{0.48\linewidth}
        \includegraphics[width=\linewidth]{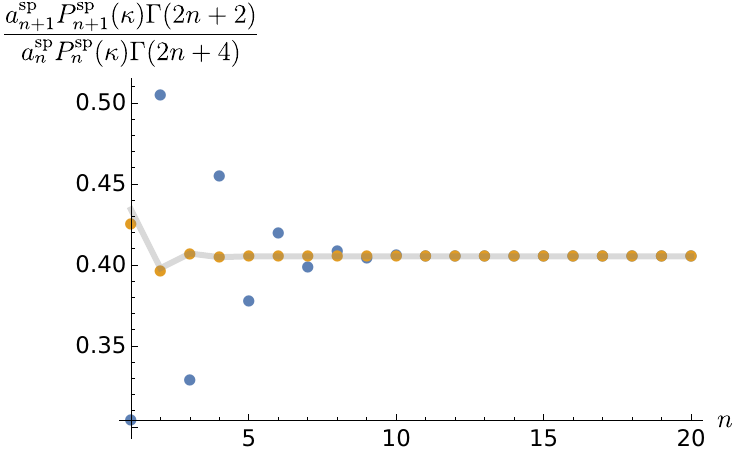}
        \caption{}
    \end{subfigure}
    \begin{subfigure}{0.48\linewidth}
        \includegraphics[width=\linewidth]{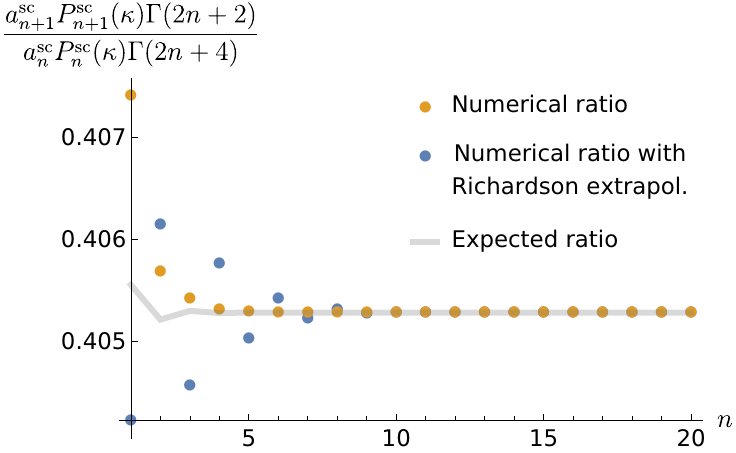}
        \caption{}
    \end{subfigure}
    \caption{Comparison of numerically computed ratios (orange) and with Richardson extrapolation applied (blue) of consecutive weak-field coefficients of (a)\, spinor and (b)\, scalar parallel fields EHLs with the expected ratio (grey) computed from the leading order contributions in Eqs.\,(\ref{eq:large-N-behaviour-spinor}) and (\ref{eq:large-N-behaviour-scalar}) respectively. The value of $\kappa$ is fixed at $\kappa=2$.}
    \label{fig:richardson}
\end{figure*}

We can also perform a similar analysis for the SQED case to get that
\begin{multline}\label{eq:large-N-behaviour-scalar}
    \frac{a_n^{\text{sc}}P_n^{\text{sc}}(\kappa)}{\Gamma(2n+2)} = 2 \kappa^2 \sum_{k=1}^{\infty} \frac{(-1)^{k-1}}{(k\pi)^{2n+3}}  \left[\kappa^{2n+1}\csch \left(\frac{k\pi}{\kappa}\right) \right. \\ \left.  + \frac{(-1)^n}{\kappa} \csch \left(k \pi \kappa \right) \right] \; ,
\end{multline}
where the coefficients $a_n^{sc}$ and $P_n^{sc}$ are defined by the asymptotic expansion
\begin{equation}\label{eq:weak-field-expansion-scalar}
    \mathcal{L}_{\text{SQED}}^{(1)} (\eta, \kappa) \sim \frac{\eta^4}{16\pi^2} \sum_{n=0}^{\infty} a_n^{sc} \ P_n^{sc}(\kappa) \ \eta^{2n}
\end{equation}
Fig.\,(\ref{fig:large-n-behaviour}) compares the growth of the $50^{\text{th}}$ weak-field coefficient of the one-loop parallel fields spinor and scalar EHL with the predicted asymptotic behaviours in Eqs.\,(\ref{eq:large-N-behaviour-spinor}) and (\ref{eq:large-N-behaviour-scalar}) as a function of $\kappa$. The coefficients' growth with $\kappa$ follows the predicted curve accurately for both the spinor and scalar cases. Fig.\,(\ref{fig:richardson}) demonstrates that numerically computed ratios of consecutive weak-field coefficients converge quickly to the expected ratios calculated from the leading order contributions to the weak-field coefficients in Eqs.\,(\ref{eq:large-N-behaviour-spinor}) and (\ref{eq:large-N-behaviour-scalar}) for both spinor and scalar EHLs. Note that the Richardson extrapolated ratios of weak-field coefficients converge much more slowly than the unextrapolated ratios, which is unlike the result that was seen for similarly parametrised expansions in similar recent studies\,\cite{DunneHarrisInhom}. This is because the large-N expansions in Eqs.\,(\ref{eq:large-N-behaviour-spinor}) and (\ref{eq:large-N-behaviour-scalar}) don't have slowly converging ``power-law" corrections but only have exponential corrections that converge much faster than power-law corrections.

\section{Resummation analysis of the parallel fields SQED and QED Euler-Heisenberg Lagrangian}
\label{sec:resum}

The Gevrey-1 divergence observed in the large-order behaviour of the coefficients of the weak-field expansions for the parallel fields Euler-Heisenberg Lagrangians (EHL) in QED and SQED, as described in Eqs.\,(\ref{eq:large-N-behaviour-spinor}) and (\ref{eq:large-N-behaviour-scalar}), indicates that these expansions are asymptotic series. Consequently, appropriate resummation techniques must be applied to extract finite results from these weak-field expansions. 

A widely used resummation approach in the context of quantum field theories is the Pad\'{e}-Borel and Pad\'{e}-Conformal-Borel summation techniques\,\cite{florio,dunne-harris-higher-loop,Dunne:1999uy,jentschura_asymptotic_2000,costin_conformal_2021,conformal-qcd}. However, most of the existing work in the literature focuses on resummation methods defined for single-variable expansions. In this section, we aim to generalise the Pad\'{e}-Borel and Pad\'{e}-Conformal-Borel summation methods to the case of a two-variable expansion by converting the two-variable expansion into a parametrised single-variable expansion as discussed in Sec.\,\ref{sec:large-n-expansion}. Furthermore, we will use both of these methods to resum the weak-field expansions of spinor and scalar QED and compare the effectiveness of the two methods.

\subsection{Pad\'{e}-Borel Resummation}\label{sec:pade-borel}
Let us start by discussing the direct application of the Pad\'{e}-Borel reconstruction. Eqs.\,(\ref{eq:weak-field-expansion-spinor}) and (\ref{eq:weak-field-expansion-scalar}) describe how we can rewrite an expansion for the parallel fields EHL in two variables $\eta$ and $\epsilon$ as an expansion in the primary variable $\eta$ and the parameter $\kappa = \epsilon/\eta$. In the resummation procedures described below, we will primarily deal with Borel transforms from the $\eta$ to the $\zeta$ plane, treating the variable $\kappa$ as a fixed parameter. Given a parametrized single-variable sum $\mathcal{S}(\eta,\kappa)$, we can define the Borel transform given by
\begin{equation}\label{eq:borel-transform-def-stripped}
    \hat{\mathcal{S}}_\lambda(\zeta,\kappa) = \sum_{n=0}^\infty \frac{a_n P_n(\kappa)}{(2n+\lambda)!} \zeta^{2n+\lambda} \; .
\end{equation}
The Borel sum is defined as a Laplace-like integral of the Borel transform such that
\begin{equation}
    \text{B}_\lambda[\mathcal{S}] (\eta,\kappa) = \eta^{-\lambda} \int_0^{\infty (1+i\epsilon)} dt \ e^{-t} \ \widehat{\mathcal{S}}_\lambda(\eta t, \kappa) \; .
\end{equation}
\begin{figure*}[!ht]
    \centering
    \begin{subfigure}{0.56\linewidth}
        \includegraphics[width=\linewidth]{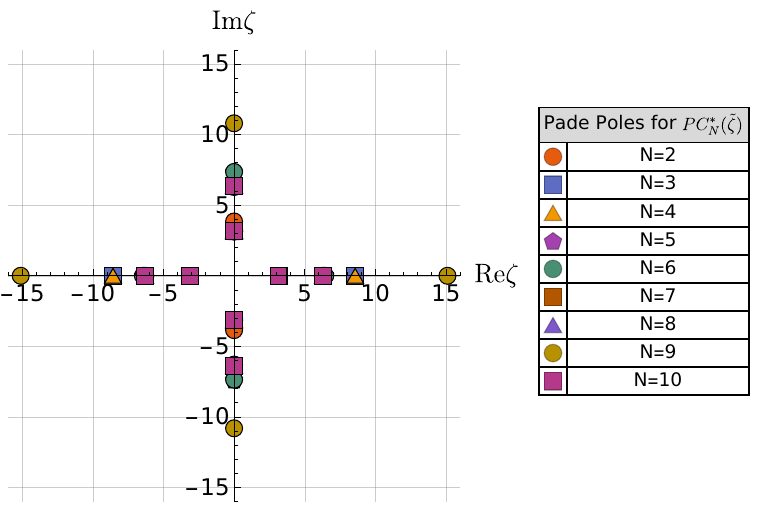}
        \caption{}
    \end{subfigure}
    \begin{subfigure}{0.4\linewidth}
        \includegraphics[width=\linewidth]{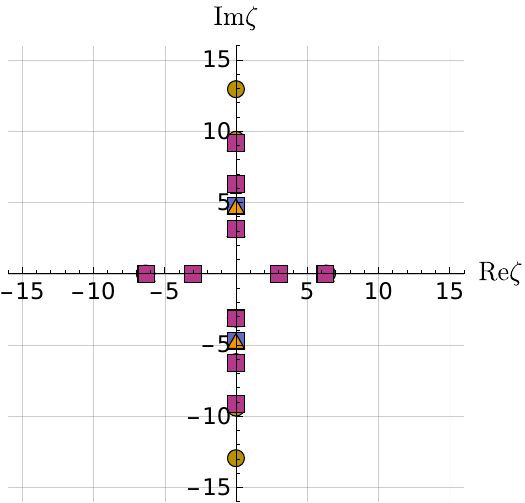}
        \caption{}
    \end{subfigure}
    \caption{The location of the poles of the Pad\'{e} sums for (a) $P^N_{N+1}\left[\widehat{\mathcal{S}}^{\text{sp}}_{\lambda,N^*} \right](\zeta,\kappa)$ and (b) $P^N_{N+1}\left[\widehat{\mathcal{S}}^{\text{sc}}_{\lambda,N^*} \right](\zeta,\kappa)$ for $\kappa=0.5$ and varying values of $N$. } 
    \label{fig:pade-borel-pole-structure}
\end{figure*}
The Borel sum $\text{B}_\lambda[\mathcal{S}]$ is asymptotic to the original perturbative series and also simultaneously encodes information about the nonperturbative behaviour of the original EHL, such as the exponentially suppressed imaginary part of the EHL, which is responsible for Schwinger pair production. However, in practical usage of the Borel reconstruction method, we often only have access to a finite number of perturbative coefficients. In such cases, the Pad\'{e}-Borel sum is defined as\,\cite{Gupta:2023gsw}
\begin{multline}\label{eq:pade-borel-def}
    \text{PB}_{N,f,\lambda} [\mathcal{S}] (\eta,\kappa) = \eta^{-\lambda} \int_0^{\infty (1+i\varepsilon)} dt \ e^{-t} \\ \times P^N_{N+f}\left[\widehat{\mathcal{S}}_{\lambda,N^*} \right] (\eta t,\kappa) \; ,
\end{multline}
where $P^N_{N+f}\left[ \hat{S}_{\lambda,N^*} \right]$ is the Pad\'{e} approximant of the Borel transform and $N^*$ is the order at which the Borel transform is truncated. Notice that the integration contour is slightly deformed away from the real axis. This deformation is important in the specific case of the parallel fields spinor and scalar EHL because of the poles on the real axis in the $\zeta$ plane, as discussed in Sec.\,\ref{sec:large-n-expansion}. This pole structure can also be seen numerically for the Borel transforms truncated at finite orders in Fig.\,(\ref{fig:pade-borel-pole-structure}), where the poles of the Pad\'{e} sum $P^N_{N+1}\left[\widehat{\mathcal{S}}_{\lambda,N^*} \right](\zeta,\kappa)$ have been plotted for the Borel transforms of the weak-field expansions of the one-loop scalar and spinor EHL's defined in Eqs.\,(\ref{eq:weak-field-expansion-spinor}) and (\ref{eq:weak-field-expansion-scalar}) for $\kappa=0.5$. While both the perpendicular fields and parallel fields cases have simple poles, the poles of the Pad\'{e} approximants of the parallel fields expansions are distributed along both the real and imaginary axes of the Borel plane, which is unlike the one-loop EHLs in the perpendicular fields case for spinor and scalar QED\,\cite{dunne-harris-higher-loop,Gupta:2023gsw} where the poles were only along the imaginary axis of the Borel plane. This is consistent with the pure electric/magnetic field limit since all the poles on the real axis situated at $k\pi/\kappa$ get pushed either to the origin or infinity in the limits $\kappa \to \infty$ or $\kappa \to 0$ respectively. As we saw in Sec.\,\ref{sec:large-n-expansion}, these poles on the real axis are responsible for the imaginary part of the one-loop EHL of parallel fields given in Eqs.\,(\ref{eq:ehl-imaginary-instanton-expansion-qed}) and (\ref{eq:ehl-imaginary-instanton-expansion-sqed}). In this section, we will compare the numerically reconstructed exponentially suppressed contributions to the spinor and scalar EHLs using the Pad\'{e}-Borel resummation method defined in Eq.\,(\ref{eq:pade-borel-def}) on a finite number of perturbative coefficients with known instanton expansions of the imaginary parts of the EHLs. We will also compare similarly reconstructed real parts with the known special function expansions in Eqs.\,(\ref{eq:ehl-numerical-real-part-qed}) and (\ref{eq:ehl-numerical-real-part-sqed}).

The contour deformation parameter $\varepsilon$ is treated as an adjustable parameter in the numerical implementation of the Pad\'{e}-Borel reconstruction, allowing optimisation for the best results. Additionally, the parameters $\lambda,f$ can also be varied to obtain better results. In this work, we set $\lambda = 1$ and $f = 1$, as these values have been demonstrated to yield optimal results for one-loop perpendicular fields spinor and scalar QED\,\cite{dunne-harris-higher-loop,Gupta:2023gsw}.

\begin{figure*}[!ht]
    \centering
    \begin{subfigure}{0.58\textwidth}
        \includegraphics[width=\textwidth]{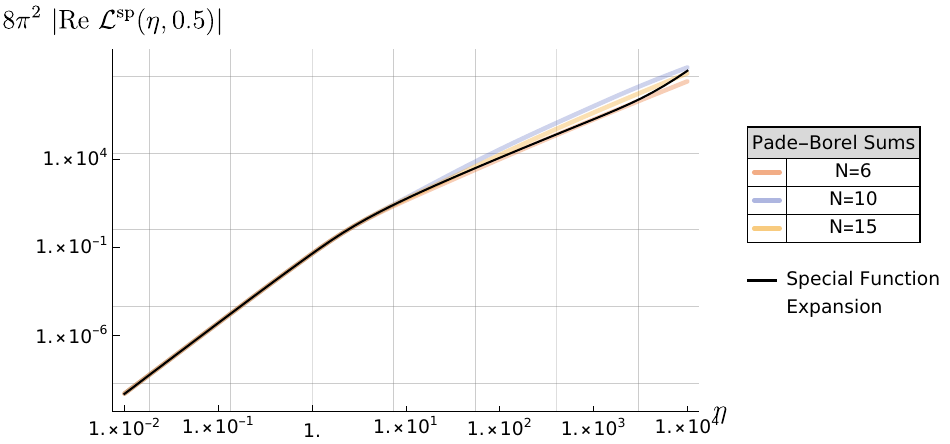} 
        \caption{}
    \end{subfigure}
    \begin{subfigure}{0.41\textwidth}
        \includegraphics[width=\textwidth]{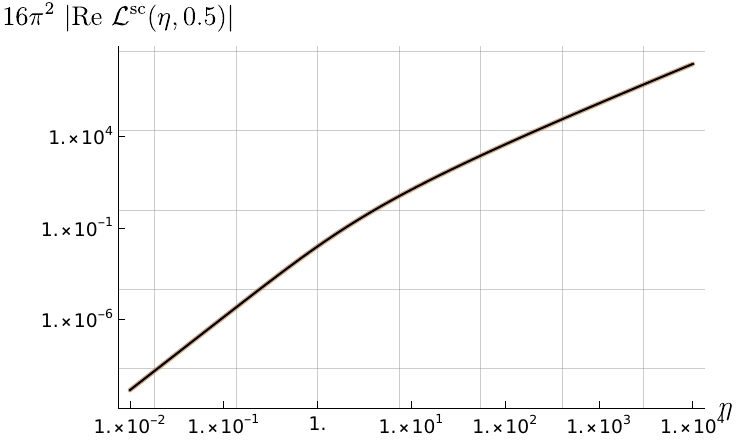}
        \caption{}
    \end{subfigure}
    \begin{subfigure}{0.48\textwidth}
        \includegraphics[width=\textwidth]{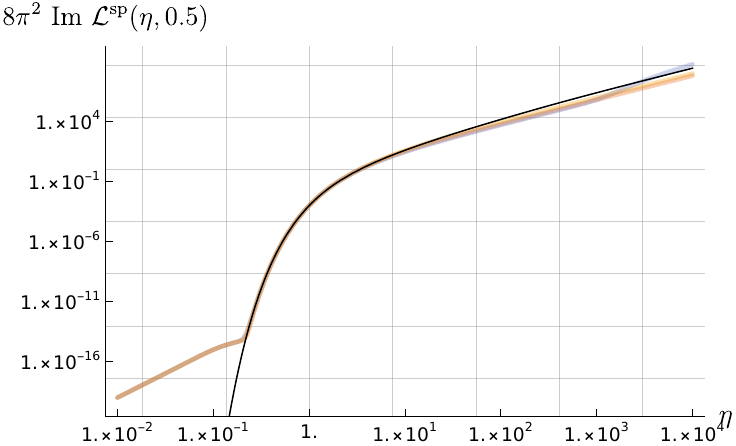}
        \caption{}
    \end{subfigure}
    \begin{subfigure}{0.48\textwidth}
        \includegraphics[width=\textwidth]{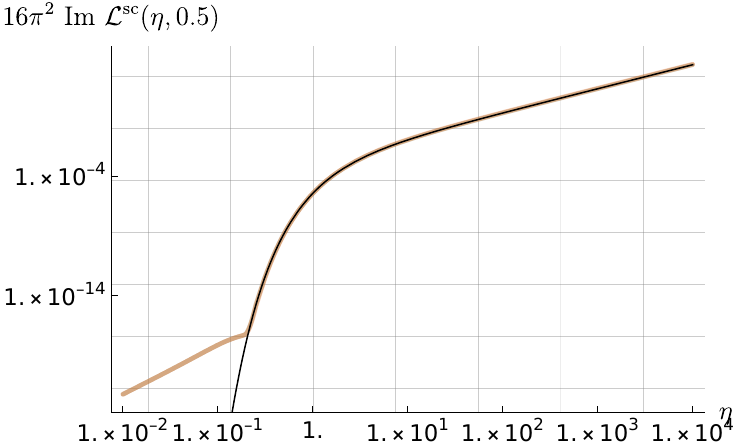}
        \caption{}
    \end{subfigure}
    \caption{The Pad\'{e}-Borel reconstruction of the one-loop parallel fields spinor and scalar EHL for different values of $N$ with $\kappa=0.5$. The parameter values used here are $\lambda=1,f=1,\varepsilon=0.1$. (a), (b) are plots of absolute values of the real part of the Pad\'{e}-Borel reconstructed spinor and scalar EHLs, respectively, compared with their special function expansions in Eqs.\,(\ref{eq:ehl-numerical-real-part-qed}) and (\ref{eq:ehl-numerical-real-part-sqed}). (c), (d) are the plots of the imaginary part of the Pad\'{e}-Borel reconstructed spinor and scalar EHLs, respectively, compared to the instanton expansions in Eqs.\,(\ref{eq:ehl-imaginary-instanton-expansion-qed}) and (\ref{eq:ehl-imaginary-instanton-expansion-sqed}).}
    \label{fig:pade-borel-x=0.5}
\end{figure*}

The results of the numerical implementation of the Pad\'{e}-Borel sum of the weak-field expansion of the parallel fields spinor and scalar EHLs for $\kappa=0.5$ and different values of $N$ are given in Fig.\,(\ref{fig:pade-borel-x=0.5}). The values of the parameters used are $\lambda=1,f=1,\varepsilon=0.1$. The real part of the Pad\'{e}-Borel method seems to match the special function expansions quite well, even for orders of $\eta \sim 10^4$ in the cases of both spinor and scalar EHLs. Although the imaginary part of the Pad\'{e}-Borel sums deviates from the instanton expansion for small values of $\eta$, when the values of the reconstructed functions are of order $\sim 10^{-15}$ (which is to be expected because the machine precision of \texttt{Mathematica} is around 16 digits), the Pad\'{e}-Borel resummation method is effective at reconstructing exponentially small imaginary contributions in the range of $\eta \sim 0.1$ to $\eta \sim 10^4$ both in the cases of scalar and spinor EHLs.

However, the Pad\'{e}-Borel resummation method is unable to perform similarly well when applied to the EHLs for higher values of $\kappa$, as shown in Figs.\,(\ref{fig:pade-conformal-borel-x=1}) and (\ref{fig:pade-conformal-borel-x=10}). For relatively larger values of $\kappa$ such as $\kappa=1$ and $\kappa=10$, the real part of the Pad\'{e}-Borel resummation values deviates significantly from the values computed from the special function expansions (given by the black line). For $\kappa=1$, the deviation of the absolute value of the real part of the Pad\'{e}-Borel from the special function expansion is of order $\sim10^4$ for $\eta \sim 10^3$, as shown in Fig.\,(\ref{fig:pade-conformal-borel-x=1}). For $\kappa=10$, the absolute value of the Pad\'{e}-Borel sum of the QED EHL seems to follow the special value expansion, but is unable to mimic the second kink in the black line at $\eta \sim 10^3$ as shown in Fig.\,(\ref{fig:pade-conformal-borel-x=10}). Since the plotted values represent absolute values, these kinks indicate changes in the sign of the EHLs and are thus critical features that resummation procedures must capture accurately. To address these limitations of the Pad\'{e}-Borel resummation procedure, we will attempt to modify our approach by using the Pad\'{e}-Conformal-Borel resummation method.

\begin{figure*}[!ht]
    \centering
    \begin{subfigure}{0.58\linewidth}
        \includegraphics[width=\linewidth]{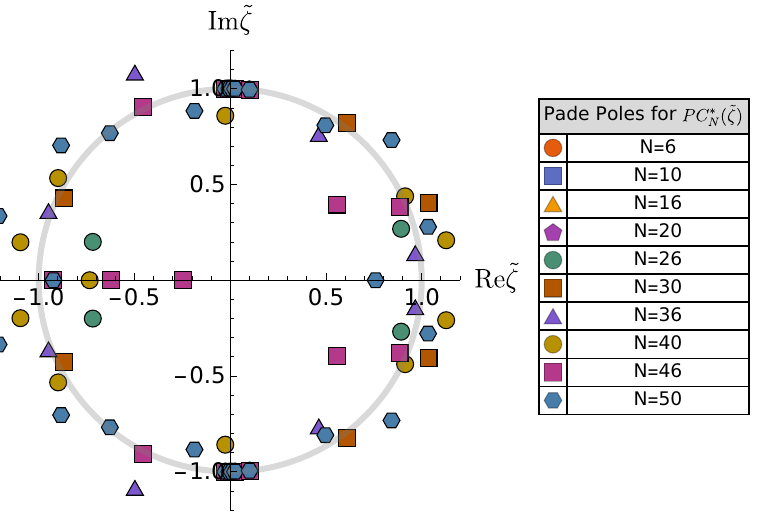}
        \caption{}
    \end{subfigure}
    \begin{subfigure}{0.41\linewidth}
        \includegraphics[width=\linewidth]{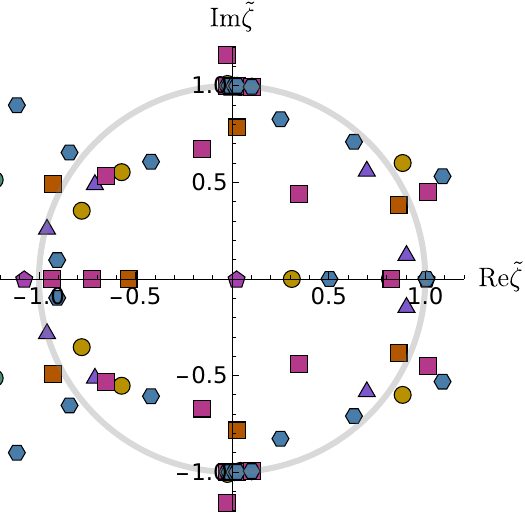}
        \caption{}
    \end{subfigure}
    \caption{The poles of the Pad\'{e} approximants $\text{PC}^*_N(\tilde{\zeta})$ in the conformal plane for (a)\,spinor, and (b)\,scalar EHLs for varying values of $N$ and $\kappa=2$.}
    \label{fig:conformal-poles}
\end{figure*}

\subsection{Pad\'{e}-Conformal-Borel Resummation}\label{sec:pade-conformal-borel}

Recent studies in resurgence and resummation techniques have extensively employed the conformal transform technique to enhance the convergence of Borel sums\,\cite{dunne-harris-higher-loop,costin_conformal_2021,conformal-qcd,jentschura_resummation_2001}. In a nutshell, the enhanced convergence achieved through conformal mappings arises from their ability to map all poles of the Borel transform onto a unit circle around the origin in the conformal plane, thus improving the ability of the Pad\'{e} approximant to capture the pole structure of the Borel transform far away from the origin.

Given the Borel transform in Eq.\,(\ref{eq:borel-transform-def-stripped}) for $\lambda =1$, we will define a new variable $\tilde{\zeta}$ as a conformal transform of the variable $\zeta$ in the Borel plane given by (see Appendix.\,\ref{app:pcbr} for more details)
\begin{equation}\label{eq:conformal-transform-definition}
    \zeta  =  \begin{cases}
         \sqrt{\frac{2 \tilde{\zeta}}{1+\tilde{\zeta}^2}} \; , & \kappa < 1  \; ,\\
        \frac{1}{\kappa} \sqrt{\frac{2 \tilde{\zeta}}{1+\tilde{\zeta}^2}} \; , & \kappa \geq 1 \;.
    \end{cases}
\end{equation}
We have introduced a factor of $1/\kappa$ in the definition for $\kappa > 1$ to ensure that the poles of the Borel transform lie outside the unit radius in the Borel plane. This adjustment is necessary because conformal transformations tend only to map poles located outside the unit circle in the Borel plane onto the circumference of the unit circle in the conformal plane. In the following analysis, we will only consider the case when $\kappa >1$ because, as we saw in the last section, the performance of the Pad\'{e}-Borel transform is reasonably good for small values of $\kappa$. 

\begin{figure*}[!ht]
    \centering
    \begin{subfigure}{0.48\textwidth}
        \includegraphics[width=\textwidth]{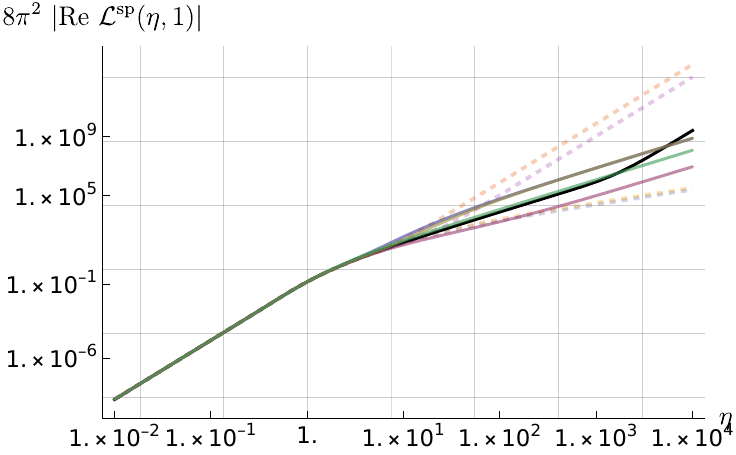} 
        \caption{}
    \end{subfigure}
    \begin{subfigure}{0.48\textwidth}
        \includegraphics[width=\textwidth]{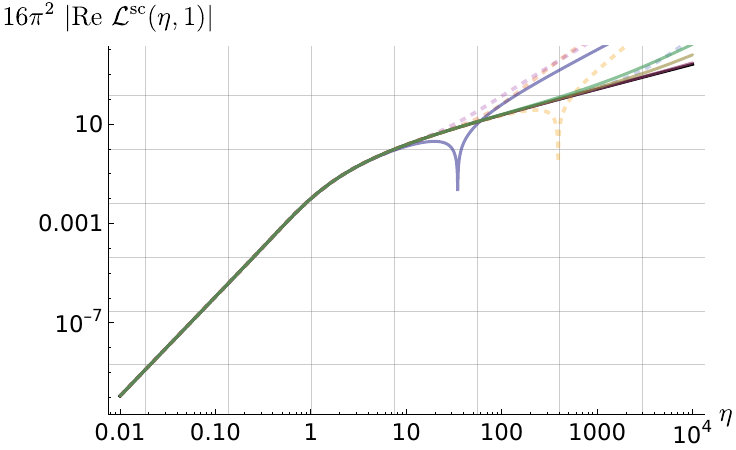}
        \caption{}
    \end{subfigure}
    \begin{subfigure}{0.58\textwidth}
        \includegraphics[width=\textwidth]{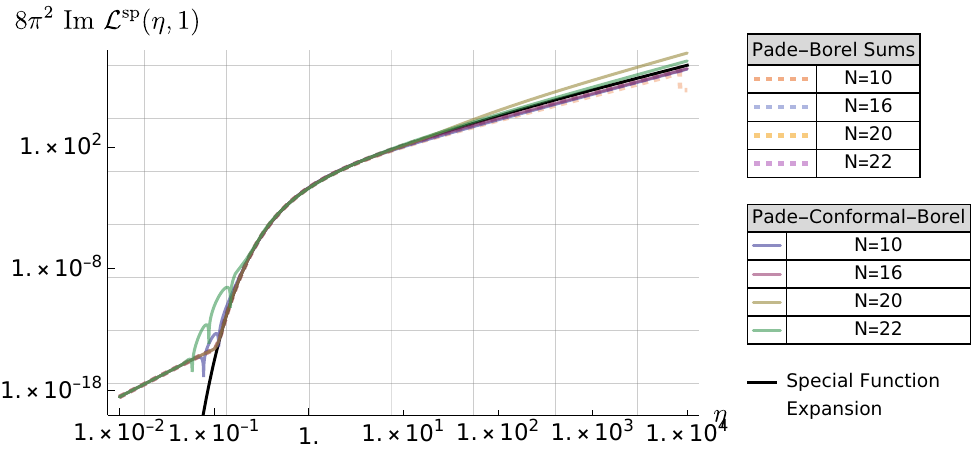}
        \caption{}
    \end{subfigure}
    \begin{subfigure}{0.41\textwidth}
        \includegraphics[width=\textwidth]{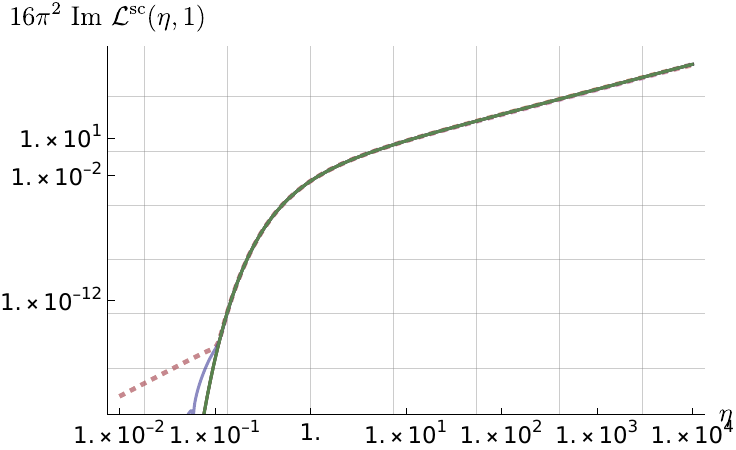}
        \caption{}
    \end{subfigure}
    \caption{The Pad\'{e}-Borel (dotted lines) and Pad\'{e}-Conformal-Borel (solid lines) reconstruction of one-loop parallel fields spinor and scalar EHLs for different values of $N$ with $\kappa=1$. (a), (b) are plots of absolute values of the real part of the Pad\'{e}-Borel reconstructed spinor and scalar EHLs, respectively, compared with their special function expansions, and (c), (d) are the plots of the imaginary part of the Pad\'{e}-Borel reconstructed spinor and scalar EHLs, respectively, compared to the instanton expansions.}
    \label{fig:pade-conformal-borel-x=1}
\end{figure*}
\begin{figure*}[!ht]
    \centering
    \begin{subfigure}{0.58\linewidth}
        \includegraphics[width=\linewidth]{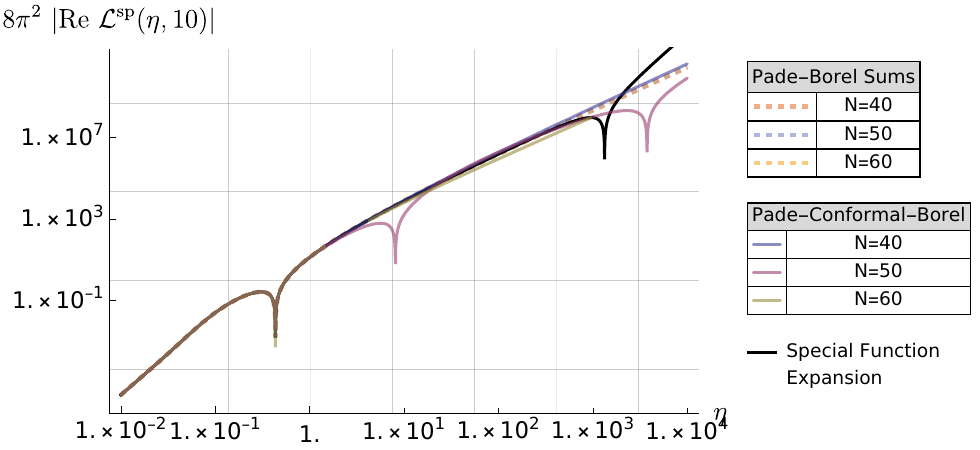}
        \caption{}
    \end{subfigure}
    \begin{subfigure}{0.41\linewidth}
        \includegraphics[width=\linewidth]{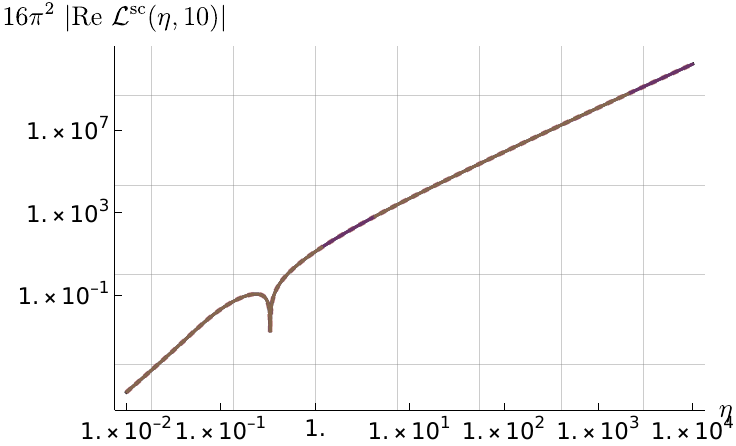}
        \caption{}
    \end{subfigure}
    \caption{The Pad\'{e}-Borel (dotted lines) and Pad\'{e}-Conformal-Borel (solid lines) reconstruction of absolute values of the real parts of (a)\,spinor and (b)\,scalar EHLs for different values of $N$ with $\kappa=10$.}
    \label{fig:pade-conformal-borel-x=10}
\end{figure*}

To avoid fractional powers of $\tilde{\zeta}$ when expressing the Borel transform in terms of the conformally-transformed variable, we take a factor of $\zeta$ out of the sum before substituting $\zeta$ in terms of $\tilde{\zeta}$ in Eq.\,(\ref{eq:borel-transform-def-stripped}). Therefore, we get the conformal Borel sum given by
\begin{equation}
    \text{C}[\widehat{\mathcal{S}}] (\tilde{\zeta},\kappa) = \sum_{n=0}^{\infty} \tilde{a}_n \tilde{P}_n(\kappa) \tilde{\zeta}^{n} \; ,
\end{equation}
where $\tilde{a}_n$ and $\tilde{P}_n(\kappa)$ are the expansion coefficients of the Borel transform in the conformal plane. Subsequently, a Pad\'{e} approximant is constructed using this conformal transform. To further improve convergence, we will introduce a factor of $1/(1+\tilde{\zeta}^2)$ to introduce simple poles at $\tilde{\zeta}=\pm i$, which correspond to the poles at infinities in the Borel plane (see Appendix\,\ref{app:pcbr} for more details). Therefore, the Pad\'{e} approximant is given by 
\begin{equation}
    P^{\lfloor N/2 \rfloor}_{\lfloor N/2 \rfloor+1}\left[\frac{1}{1+\tilde{\zeta}^2} \ \text{C} [\widehat{\mathcal{S}}](\tilde{\zeta},\kappa) \right] = \text{PC}^*_N [\widehat{\mathcal{S}}](\tilde{\zeta},\kappa) \; ,
\end{equation}
where $\lfloor N/2 \rfloor$ denotes the floor function of $N/2$. The poles of the Pad\'{e} transform $P_N^*(\tilde{\zeta})$ in the conformal plane are plotted in Fig.\,(\ref{fig:conformal-poles}) --- it is numerically apparent that most poles in the conformal plane are accumulated near the unit circle in the conformal plane. There is a high density of poles near $\tilde{\zeta} = \pm i$, which is the accumulation point of the poles far away from the origin in the Borel plane.

The Pad\'{e}-Conformal-Borel sum can be found by taking a Laplace-like transform of the modified Pad\'{e} approximant $\text{PC}^*_N$ such that
\newline
\begin{multline}
    \text{PCB}_{N} \left[ \mathcal{S} \right] (\eta,\kappa) = \eta^{-1} \int_0^{\infty} dt \  e^{-t} \eta t \ (1+\widetilde{\zeta}^2(\eta t)) \\ \times \text{PC}^*_N [ \widehat{\mathcal{S}}]  ( \widetilde{\zeta}(\eta t),\kappa) \; ,
    \end{multline}
where $\tilde{\zeta}(\eta t)$ is defined implicity as function of the integration variable via Eq.\,(\ref{eq:conformal-transform-definition}) which maps the conformal variable $\tilde{\zeta}$ to the Borel plane variable $\zeta = \eta t$.

Using these definitions, the Pad\'{e}-Conformal-Borel transforms were computed for the QED and SQED EHL for $\kappa=1$ and $\kappa=10$ and compared with the corresponding special function expansions in Eq.\,(\ref{eq:ehl-numerical-real-part-qed}), (\ref{eq:ehl-numerical-real-part-sqed}), (\ref{eq:ehl-imaginary-instanton-expansion-qed}), and (\ref{eq:ehl-imaginary-instanton-expansion-sqed}) and also the Pad\'{e}-Borel transforms defined in Sec.\,\ref{sec:pade-borel} in Figs.\,(\ref{fig:pade-conformal-borel-x=1}) and (\ref{fig:pade-conformal-borel-x=10}). For $\kappa=1$, it is apparent that the Pad\'{e}-Conformal-Borel transform does a much better job at reconstructing the real part of the underlying function than the simple Pad\'{e}-Borel transform both in the cases of the QED and SQED EHLs--- the Pad\'{e}-Conformal-Borel resummation exhibits an improvement of $\sim 10^4$ as compared to the Pad\'{e}-Borel sum at $\eta \sim 10^3$ in the case of QED and that of $\sim 10^3$ at $\eta \sim 10^4$ in the case of SQED at expansion order $N=22$. The performance of both Pad\'{e}-Borel and Pad\'{e}-Conformal-Borel transforms in reconstructing the imaginary parts of the EHLs is similar. Although the Pad\'{e}-Borel transform follows the special value expansion of the real part of the spinor EHL more closely for $\kappa=10$, the Pad\'{e}-Conformal-Borel transform is more effective at capturing both the kinks observed in the plot of Fig.\,(\ref{fig:pade-conformal-borel-x=10}). Therefore, the Pad\'{e}-Conformal-Borel transform is better at capturing the qualitative behaviour of the EHL in nonperturbative regimes of field strengths of order $\sim 10^4$ using finite number of perturbative coefficients--- as noted above, this improvement in performance may be attributed to the ability of conformal transforms to map the poles in the Borel plane onto compact spaces in the conformal plane, thereby enhancing the ability of the Pad\'{e} approximants to resolve the pole structure of the theory. The conformal improvement in the case of scalar EHLs is not as remarked; the Pad\'{e}-Borel and Pad\'{e}-Conformal-Borel methods have similar performances for $\kappa=10$.

\section{Conclusions}
\label{sec:conclusion}

In this work, we have presented a systematic and comprehensive resurgent analysis of the one-loop Euler–Heisenberg Lagrangian in quantum electrodynamics and scalar quantum electrodynamics in the most general constant field configuration---where both electric and magnetic fields are simultaneously present. While previous resurgent investigations of the EHL have focused almost exclusively on perpendicular fields limits—purely magnetic or purely electric backgrounds—the parallel fields configuration studied here introduces new analytic features and provides a natural completion of the resurgence programme for constant electromagnetic backgrounds\,\cite{Dunne:1999uy,florio,DunneHarrisInhom,dunne-harris-higher-loop,Gupta:2023gsw}.

Reorganising the double weak-field expansion into a single asymptotic series with the ratio of field invariants held fixed, and by employing Borel–dispersion techniques, we obtained closed asymptotic expressions for the coefficients. Thereby, we give an explicit derivation of the large-order behaviour of the weak-field expansion coefficients of the parallel fields EHL in both spinor and scalar QED. These formulas reveal a nontrivial interplay between alternating and non-alternating factorial growth, with the relative dominance controlled by the electric-to-magnetic field ratio. The resulting structure interpolates smoothly between the known perpendicular fields magnetic and electric limits, reproducing them precisely in the corresponding regimes. Notably, the large-order growth encodes the full instanton content governing Schwinger pair production, thereby establishing a precise resurgence relation between perturbative weak-field data and nonperturbative vacuum decay rates in the parallel fields setting.

Another inference from the analyses is the identification of a fundamentally new Borel-plane structure in the parallel fields problem. Unlike the perpendicular fields case, where singularities lie exclusively along one axis, the parallel fields EHL exhibits families of Borel singularities on both the real and imaginary axes. These singularities are directly tied to the coexistence of electric and magnetic backgrounds and are responsible for the simultaneous presence of alternating and non-alternating contributions in the weak-field series. This richer singularity structure, in turn, underlies many of the distinctive resummation features observed in the functional reconstructions.

Building on the analytic large-order results, we investigated Pad\'{e}–Borel resummation techniques adapted to the unique structure of the problem and demonstrated that, using only a finite number of weak-field coefficients, it is possible to reconstruct both the real and imaginary parts of the parallel fields EHL with high accuracy over a broad range of field strengths and field ratios. For moderate values of the electric-to-magnetic ratio, the Pad\'{e}–Borel approach successfully reproduces the exact special-function representations of the EHL, including the exponentially suppressed imaginary part associated with Schwinger pair production.

For larger field ratios, particularly in (spinor) QED, we found that the standard Pad\'{e}–Borel method becomes insufficient to capture subtle strong-field features, most notably sign-changing behaviour in the real part of the effective Lagrangian. To explore these limitations, we explored a Pad\'{e}–Conformal–Borel resummation scheme tailored to the parallel fields Borel-plane geometry. The conformal technique and improvement lead to a marked enhancement in convergence and accuracy for QED, allowing the resummed expressions to faithfully reproduce nontrivial strong-field structures that are missed by the conventional approach. Similar conformal mappings also lead to minor improvements in the reconstruction of the SQED EHL.

Several natural directions for future work emerge from this study. A significant extension is the application of the present framework to higher-loop corrections, where the Borel singularities are expected to develop branch-cut structures associated with virtual particle interactions\,\cite{dunne-harris-higher-loop}, leading to genuinely multi-parameter trans-series. Another promising avenue is the extension to inhomogeneous or time-dependent electromagnetic backgrounds\,\cite{DunneHarrisInhom}, where additional scales are present, and new resurgent phenomena are anticipated. Finally, the techniques developed here may find direct application in strong-field QED processes\,\cite{non-linear-trident-resummation,Torgrimsson:2022ndq,Torgrimsson:2021wcj,Torgrimsson:2021zob} relevant to upcoming high-intensity laser experiments, where mixed electric–magnetic configurations and nonperturbative effects play a central role.

In summary, this work develops the theoretical exploration of the parallel fields Euler–Heisenberg Lagrangian in  QED and SQED as a natural and necessary completion of the resurgence and resummation programme for constant electromagnetic backgrounds, providing both conceptual insights and potential practical tools for reconstructing nonperturbative physics from finite perturbative data.

\section*{Aknowledgements}
DG acknowledges support from a graduate student research assistantship from the University of Illinois, Urbana-Champaign. Parts of this work were completed during the conference ``Gravity 2025: New Horizon of Black Hole Physics'', at the Yukawa Institute for Theoretical Physics, Kyoto University, and AT acknowledges discussions there and thanks the organisers for their hospitality. Some of the numerical computations presented in this work were performed using \texttt{Mathematica}.
\appendix 

\section{Pad\'{e} Conformal Borel Resummation}\label{app:pcbr}
The method of Pad\'{e}-Conformal-Borel resummation is a well-known technique\,\cite{dunne-harris-higher-loop,CALICETI20071,costin_conformal_2021,Costin:2019xql} to improve the convergence of Borel sums by mapping the poles in the Borel plane onto a compact space by a conformal transformation. In this work, we will use conformal transformations of the form
\begin{equation}\label{appeq:conformal-transform-def}
    \tilde{\zeta} = \frac{1-\sqrt{1-\zeta^4}}{\zeta^2}
\end{equation}
where $\tilde{\zeta}$ is the conformal variable. The most interesting and useful feature of this conformal transform is that it maps points on the real and imaginary axes with absolute values greater than 1 onto a unit circle in the conformal plane. To see this, suppose that the point $\zeta_0$ in the Borel plane is either on the real or imaginary axis with absolute value bigger than 1. Then, $\sqrt{1-\zeta_0^4}$ is purely imaginary. So, we have that
\begin{equation}
    \tilde{\zeta}_0 = \frac{1 + i \sqrt{\zeta_0^4 -1}}{\zeta_0^2}
\end{equation}
where $\tilde{\zeta}_0$ is the conformal transform of $\zeta_0$. Therefore, the absolute value of $\tilde{\zeta}_0$ is given by
\begin{equation}
\begin{split}
    |\tilde{\zeta}_0|^2 &= \frac{1+\zeta_0^4-1}{\zeta_0^4}  \\
    &=1
\end{split}
\end{equation}
Furthermore, notice the asymptotic behaviour of the conformal mapping: as the Borel variable approaches infinity along the real axis ($\zeta \to \pm \infty$), the conformal variable $\tilde{\zeta}$ tends toward $-i$. Similarly, in the limit $\zeta \to \pm i \infty$ along the imaginary axis, $\tilde{\zeta} \to i$. As established in Sec.\,(\ref{sec:large-n-expansion}), the poles of the QED and SQED Euler-Heisenberg Lagrangians are situated along the real and imaginary axes of the Borel plane. Consequently, implementing a conformal transformation prior to Pad\'{e} approximation enhances the ability of the Pad\'{e} functions to accurately capture singularities located far from the origin by mapping them onto a compact space.

Given the Borel transform of the form
\begin{equation}
    \widehat{\mathcal{S}} (\zeta) = \sum_{n=0}^{\infty} \frac{a_n \zeta^{2n}}{\Gamma(2n+2)}
\end{equation}
the conformal transform is found by substituting the the Borel variable $\zeta$ in terms of the conformal variable $\tilde{\zeta}$. For the conformal transform given in Eq.\,(\ref{appeq:conformal-transform-def}), we have that
\begin{equation}
   \zeta = \sqrt{\frac{2\tilde{\zeta}}{1+\tilde{\zeta}^2}} 
\end{equation}
So, the conformal transform of the Borel transform in this case is defined as
\begin{equation}
    \text{C}[\widehat{\mathcal{S}}] (\tilde{\zeta}) = \sum_{n=0}^{\infty} \frac{a_n}{\Gamma(2n+2)} \left(\sqrt{\frac{2 \tilde{\zeta}}{1+\tilde{\zeta}^2}} \right)^{2n} = \sum_{n=0}^{\infty} \tilde{a}_n \tilde{\zeta}^n
\end{equation}
The Pad\'{e} approximant of this conformal transform is then given by
\begin{equation}
    \text{PC}_N[ \widehat{\mathcal{S}}] (\tilde{\zeta}) = P^{\lfloor N/2 \rfloor}_{\lfloor N/2 \rfloor + 1} \left[\text{C} [\widehat{\mathcal{S}}] \right] (\tilde{\zeta})
\end{equation}
The symbol $P^{\lfloor N/2 \rfloor}_{\lfloor N/2\rfloor +1}$ denotes a Pad\'{e} approximant with polynomial of degree $\lfloor N/2 \rfloor$ (floor of $N/2$) in the numerator and polynomial of degree $\lfloor N/2 \rfloor + 1 $ in the denominator. In order to recover a function that is asymptotic to the original function, we perform a Laplace-like transform to get the Pad\'{e}-Conformal-Borel sum such that
\begin{equation}
    \text{PCB}_N[\mathcal{S}]  (\eta) = \eta^{-1} \int_0^{\infty (1+i\epsilon)} dt \ e^{-t/\eta} \ t\ \text{PC}_N [\widehat{\mathcal{S}}] (\tilde{\zeta}(t))
\end{equation}
where $\tilde{\zeta}(t)$ is the conformal transformation function in Eq.\,(\ref{appeq:conformal-transform-def}) that relates the Borel plane variable $t$ and the conformal plane variable $\tilde{\zeta}$. 

This is the basic form of the Pad\'{e}-Conformal-Borel transform that we use in Sec.\,\ref{sec:pade-conformal-borel}. Additionally, in order to better capture the poles at real and imaginary infinities in the Borel plane that get mapped to $\pm i$ in the conformal plane, we also introduce a modified Pad\'{e} approximant in the conformal plane defined as
\begin{equation}
    \text{PC}^*_N[\widehat{\mathcal{S}}] (\tilde{\zeta}) = P^{\lfloor N/2 \rfloor}_{\lfloor N/2 \rfloor+1} \left[\frac{1}{\tilde{\zeta}^2 + 1}\text{B} [\widehat{\mathcal{S}}] (\tilde{\zeta}) \right] 
\end{equation}
Then, the modified Pad\'{e}-Conformal-Borel sum is given by
\begin{multline}
    \text{PCB}^*_N [\widehat{\mathcal{S}}] (\eta) = \eta^{-1} \int_0^{\infty(1+i\epsilon)} dt \ e^{-t/\eta} \ t \ (\tilde{\zeta}^2(t)+1) \\ \times \text{PC}^*_N [\widehat{\mathcal{S}}](\tilde{\zeta}(t))
\end{multline}

\bibliography{TwoFieldEHLResurgence}

\end{document}